\pdfoutput=1
\documentclass[a4paper,12pt]{article}
\usepackage{amsfonts}
\usepackage{mathrsfs}
\usepackage{amsmath}
\usepackage{amssymb}
\usepackage{framed}
\usepackage[medium]{titlesec}
\usepackage{bm}
\usepackage{cite}
\usepackage[normalem]{ulem}
\usepackage{extarrows}
\usepackage{slashed}
\usepackage{isodateo}
\usepackage{graphicx}
\usepackage{xcolor}
\usepackage[bookmarksnumbered=true,bookmarksopen=true]{hyperref}
 \hypersetup{colorlinks,%
             linkcolor=[rgb]{0,0.3,0.6}, %
             citecolor=[rgb]{0,0.3,0.6}, %
             urlcolor=[rgb]{0,0.3,0.6}}
\usepackage[hmargin=.7in,vmargin=1.1in]{geometry}
\usepackage{indentfirst}
\usepackage{booktabs}

\graphicspath{{fig/}}

\newcommand{\vect}[1]{\mbox{\boldmath${#1}$}}
\newcommand{\lmk}{\left(}
\newcommand{\rmk}{\right)}

\newcommand{\lkk}{\left[}
\newcommand{\rkk}{\right]}
\newcommand{\lla}{\left\langle}
\newcommand{\rra}{\right\rangle}
\newcommand{\vex}{{\vect x}}
\newcommand{\veh}{{\vect h}}
\newcommand{\ven}{\vect n}

\newcommand{\veb}{{\bf e}}

\def\gpcyr{\rm \,Gpc^{-3}\,yr^{-1}}

\def\mc{{{\mathcal{M}}}_{\rm c}}

\usepackage{mdframed}

\newmdenv[skipabove=0pt,%
          skipbelow=5pt,%
          leftmargin=0pt,%
          rightmargin=0pt,%
          innertopmargin=-5pt,%
          innerbottommargin=7pt,%
          innerleftmargin=2pt,%
          innerrightmargin=2pt,%
          splittopskip=0pt,%
          splitbottomskip=0pt,%
          linewidth=0pt,%
          nobreak=true]%
          {keyeqn2}

\newmdenv[backgroundcolor=gray!15,%
          skipabove=0pt,%
          skipbelow=5pt,%
          leftmargin=0pt,%
          rightmargin=0pt,%
          innertopmargin=-5pt,%
          innerbottommargin=7pt,%
          innerleftmargin=2pt,%
          innerrightmargin=2pt,%
          splittopskip=0pt,%
          splitbottomskip=0pt,%
          linewidth=0pt,%
          nobreak=true]%
          {keyeqn}

\begin{document}

\title{\Large\textbf{On networks of space-based gravitational-wave detectors}\\[2mm]}

\author{Rong-Gen Cai$^{1,2,3}$,
Zong-Kuan Guo$^{1,3,4}$,
Bin Hu$^{5}$,
Chang Liu$^{3,4}$,\\
Youjun Lu$^{6,7}$,
Wei-Tou Ni$^{8,9}$,
Wen-Hong Ruan$^{3}$,
Naoki Seto$^{10}$,\\
Gang Wang$^{11}$,
Yue-Liang Wu$^{1,3,12}$}

\date{}
\maketitle

\begin{center}
\noindent \normalsize{${}^{1}$\emph{CAS Key Laboratory of Theoretical Physics, Institute of Theoretical Physics, Chinese Academy of Sciences, Beijing 100190, China}}\\
\normalsize{${}^{2}$\emph{School of Physical Science and Technology, Ningbo University, Ningbo 315211, China}}\\
\normalsize{${}^{3}$\emph{School of Fundamental Physics and Mathematical Sciences, Hangzhou Institute for Advanced Study, University of Chinese Academy of Sciences, Hangzhou 310024, China}}\\
\normalsize{${}^{4}$\emph{School of Physical Sciences, University of Chinese Academy of Sciences, No.19A Yuquan Road, Beijing 100049, China}}\\
\normalsize{${}^{5}$\emph{Department of Astronomy, Beijing Normal University, Beijing 100875, China}}\\
\normalsize{${}^{6}$\emph{National Astronomical Observatories, Chinese Academy of Sciences, 20A Datun Road, Beijing, 100101, China}}\\
\normalsize{${}^{7}$\emph{School of Astronomy and Space Science, University of Chinese Academy of Sciences, No. 19A Yuquan Road, Beijing, 100049, China}}\\
\normalsize{${}^{8}$\emph{International Centre for Theoretical Physics Asia-Pacific, University of Chinese Academy of Sciences, Beijing, China}}\\
\normalsize{${}^{9}$\emph{State Key Laboratory of Magnetic Resonance and Atomic and Molecular Physics, Innovation Academy for Precision Measurement Science and Technology (APM), Chinese Academy of Sciences, Wuhan 430071, China}}\\
\normalsize{${}^{10}$\emph{Department of Physics, Kyoto University, Kyoto 606-8502, Japan}}\\
\normalsize{${}^{11}$\emph{Shanghai Astronomical Observatory, Chinese Academy of Sciences, Shanghai, 200030, China}}\\
\normalsize{${}^{12}$\emph{International Centre for Theoretical Physics Asia-Pacific, Beijing/Hangzhou, China}}
\end{center}

\begin{abstract}
\vspace{2mm}
The space-based laser interferometers, LISA, Taiji and TianQin, are targeting to observe milliHz gravitational waves (GWs) in the 2030s. The joint observations from multiple space-based detectors yield significant advantages. In this work, we recap the studies and investigations for the joint space-based GW detector networks to highlight: 1) the high precision of sky localization for the massive binary black hole (BBH) coalescences and the GW sirens in the cosmological implication, 2) the effectiveness to test the parity violation in the stochastic GW background observations, 3) the efficiency of subtracting galactic foreground, 4) the improvement in stellar-mass BBH observations. We inspect alternative networks by trading off massive BBH observations and stochastic GW background observation.
\end{abstract}

\footnotetext{E-mail addresses: cairg@itp.ac.cn (R.G. Cai), guozk@itp.ac.cn (Z.K. Guo), bhu@bnu.edu.cn (B. Hu), liuchang@alumni.itp.ac.cn (C. Liu), luyj@nao.cas.cn (Y.J. Lu), weitou@gmail.com (W.T. Ni), ruanwenhong@ucas.ac.cn (W.H. Ruan), seto@tap.scphys.kyoto-u.ac.jp (N. Seto), gwang@shao.ac.cn (G. Wang), ylwu@ucas.ac.cn (Y.L. Wu)}
\newpage
\tableofcontents
\newpage

\section{Introduction}
The sky localization of gravitational-wave (GW) sources is one of key scientific tasks for GW observations.
Rapid and accurate localization is crucial to help follow-up observations of their electromagnetic counterparts.
Once their host galaxies  are uniquely identified by observations of their electromagnetic counterparts,
these GW events can be  used as standard sirens to measure the cosmological parameters~\cite{Schutz:1986gp}.
If GW events are not accompanied by electromagnet counterparts,
these GW events, as so-called dark sirens, can measure the cosmological parameters by statistical techniques using galaxy catalogues.
In this case, precision localization of GW sources reduces the number of possible host galaxies,
which improves measurements of the cosmological parameters.
However, a single laser interferometer detector is insensitive to the sky location of transient GW sources.
In principle, the location of sources can be determined by triangulation
based on the differences in times of arrival at the different detector sits in a network of more than three detectors.
Actually, including the amplitude and phase of the signal enables the source to be localized to a single sky path in a network of three detectors.

The two detectors of advanced LIGO in the United States, one in Hanford, Washington and the other  in Livingston, Louisiana,
simultaneously observed the first GW signal from the coalescence of stellar-mass binary black hole  (BBH), GW150914~\cite{LIGOScientific:2016aoc}.
During the second observing run, the advanced Virgo detector near Pisa, Italy, joined the two advanced LIGO detectors.
The GW event of GW170814 was for the first time observed with a network of three detectors~\cite{LIGOScientific:2017ycc}.
Such a network improves the sky localization of the source,
reducing the 90\% credible area on the sky from 1160 deg$^2$ using only two advanced LIGO detectors to 60 deg$^2$ with the LIGO-Virgo network.
For the signal of GW170817 from a binary neutron star inspiral,
the source was localized to a region of 190 deg$^2$ using only two advance LIGO detectors~\cite{LIGOScientific:2017vwq}.
Although the signal is not visible in the Virgo data,
adding the Virgo data allowed a precise sky localization to an area of 28 deg$^2$.
This measurement played an important role in identifying the host galaxy, NGC 4993.
The KAGRA detector located in Japan will join the LIGO-Virgo network during the fourth observing run.
The LIGO-Virgo-KAGRA network of four detectors are expected to improve the sky localization of GW sources.
LIGO-India is a planned GW detector that will be located in India as part of the global network.
Adding the LIGO-India detector to the existing LIGO-Virgo-KAGRA network will further improve the localization accuracy.

Unlike ground-based GW observations, a single space-based GW detector is able to localize the sky position of GW sources
including massive BBHs, galactic binaries (GBs) in the Milk Way and extreme mass ration inspirals.
Such sources are usually visible to the detector for several days, months, or even years.
Since the detector is moving relative to sources,
the detector can be effectively regarded as a network of multiple detectors in different locations at different times.
However, a single space-based GW detector could not accurately determine the source position on the sky.
For example, LISA, a space observatory project proposed by the European Space Agency,
achieves an angular resolution of about $0.3$ square degrees
for the coalescence of massive black hole binaries~\cite{Cutler:1997ta}.
LISA is planned to launch between 2030 and 2035 \cite{LISA:2017pwj}.
Taiji proposed by the Chinese Academy of Sciences is expected to launch during the same period as LISA.
If Taiji joins LISA, the LISA-Taiji network is expected to significantly improve the sky
localization of massive BBHs~\cite{Ruan:2020smc}.
For a BBH with a total mass of $10^5$ solar mass,
the LISA-Taiji network may improve the source localization by four orders of
magnitude compared to an individual detector~\cite{Ruan:2019tje}.
This result implies that it is possible to completely identify the host galaxy only from GW observations
prior to the merger.

Besides improving the localization accuracy, the network of detectors increases the detection rate
and improves the isotropy of the network antenna pattern~\cite{Schutz:2011tw}.
Seto pointed out that the network is able to measure the Stokes parameter of GWs, 
which characterizes the asymmetry of the amplitudes of right- and left-handed waves~\cite{Seto:2007tn}. 
Such the asymmetry is closely related to parity violation in the early Universe.
This paper presents some significant progresses on the study for the  space-based GW detector networks.
We shall focus on the sky localization of sources, test of parity violation, 
substraction of the galactic foreground, and improvement of the detection number of stellar BBHs.

The organization of this paper is as follows. In Sec.~\ref{sec:ls} we introduce the conception of space-based GW detector networks. We show that the localization accuracy of massive black hole binaries can be significantly improved by the LISA-Taiji network. This facilitates the cosmological implications such as standard sirens and dark sirens. In Sec.~\ref{sec:test_pv} we illustrate the effectiveness to test the parity violation in the stochastic GW background (SGWB). Sec.~\ref{sec:sgf} presents the galactic foreground subtraction with space-based GW detector networks. In Sec.~\ref{sec:sbbh} we present the improvement of the detection number of stellar BBHs by the LISA-Taiji network. In Sec.~\ref{sec:an} alternative networks are discussed. Sec.~\ref{sec:cp} is our conclusions and prospects.

\section{Localization of sources \label{sec:ls}}

For GW observations, fast and accurately localizing GW sources through data analysis is one of the key scientific tasks.
Currently, the second-generation ground-based laser interferometer GW detectors mainly focus on transient GW signals emitted from coalescing stellar-mass compact binaries which vary in the frequency range of $10\text-1000 \ \text{Hz}$.  
The sensitivity of the detectors to GWs is not directional, so it is hard to localize the sky position of the GW source by a single detector.
However, with a network of three or more detectors, the sky position can be inferred by triangulation and analyzing the differences in the phase and amplitude detected by each detector.
Unlike ground-based detectors, the space-based GW detectors focus on GW signals within frequency range of $10^{-4}-10^{-1} \ \text{Hz}$, which covers many types of sources~\cite{LISA:2017pwj,Hu:2017mde,ruan2020taiji,luo2016tianqin}.
In general, the GW signals can be observed by a space-based detector for extended periods of time, ranging from several days to years.
During the observation of a GW signal, the antenna pattern function of the detector varies with the motion of the detector in space, which helps to localize the sky position of the source.
\begin{figure*}[!htbp]
\centering
\includegraphics[height=10cm]{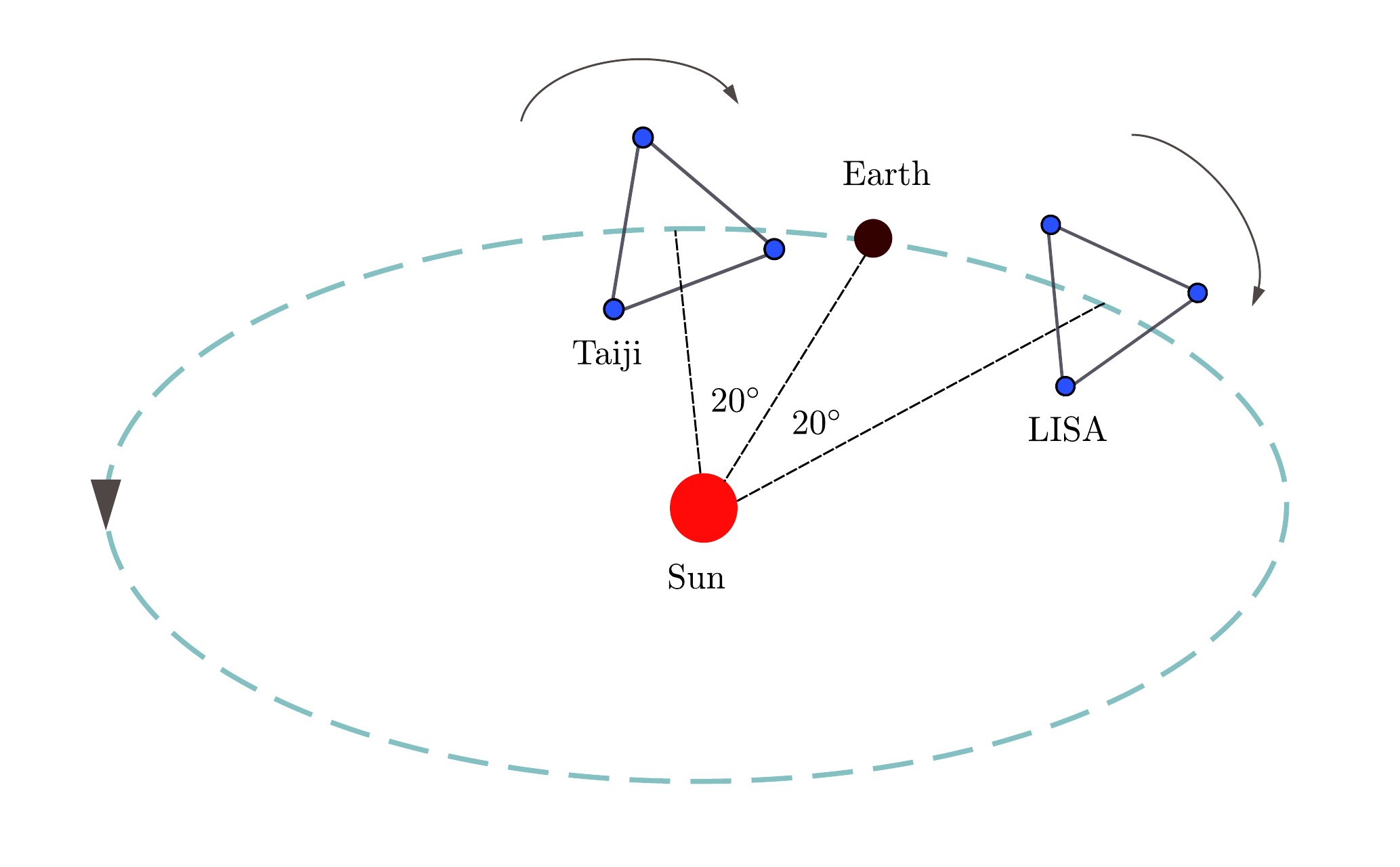}
\caption{Configuration of the LISA-Taiji network. The LISA constellation follows the Earth by $20^\circ$, while the Taiji constellation moves ahead of the Earth by $20^\circ$.}
\label{fig:LISA_Taiji}
\end{figure*}

\subsection{Localization accuracy of coalescing massive BBHs}
Particularly, the coalescing massive BBHs with total masses between $10^4 M_\odot$ and $10^8 M_\odot$ which emit GW signals accumulating large signal-to-noise ratios (SNRs) in a matter of days or weeks are primary sources of space-based GW detection~\cite{LISA:2017pwj, ruan2020taiji,hu2017science}.
The systems may also produce detectable electromagnetic emission associated with final merger~\cite{bogdanovic2022electromagnetic}.
The fast and accurate localization of these sources from GW observations is significant for uniquely identifying the host galaxy, and it allows the detailed arrangements for follow-up electromagnetic spectroscopic observations.
For a detected GW signal with a significant SNR, the errors of source parameters can be estimated by Fisher information matrix, and the elements of the matrix is given by~\cite{cutler1998angular}
\begin{equation}
    \label{eq:fisher}
    \Gamma_{ij} = \sum_{\text{det}} \left( \frac{\partial h}{\partial \lambda_i} \middle| \frac{\partial h}{\partial \lambda_j}\right)_{\text{det}},
\end{equation}
where $h$ is the GW waveform in frequncy domain and $\lambda_i$ denotes the $i\text{-th}$ source parameter. In Eq.~\ref{eq:fisher}, the summation is operated on multiple detectors and the noise-weighted inner product $(g|h)$ is given by
\begin{equation}
    \label{eq:product}
    (g|h) = 4\text{Re}\int_{0}^{\infty} \frac{g^*(f)h(f)}{S(f)}df,
\end{equation}
where $S(f)$ is the one-sided power spectral density of the detector noise.
Considering the leading order, the variance-covariance matrix of the source parameters can be obtained by~\cite{cutler1998angular}
\begin{equation}
    \label{eq:sigma_ij}
    \sigma_{ij} = \langle \Delta \lambda_i \Delta\lambda_j \rangle = (\Gamma^{-1})_{ij}.
\end{equation}
The sky location of the GW source is determined by the sky coordinates $(\theta, \phi)$ and the luminosity distance $d_L$.
The error of the sky coordinates can be converted into the angular resolution $\Delta \Omega_S$ by~\cite{Ruan:2019tje}
\begin{equation}
    \Delta \Omega_s=2\pi |{\rm sin}\theta | \sqrt{\langle \Delta \theta^2 \rangle \langle \Delta \phi^2 \rangle-\langle \Delta \theta \Delta \phi \rangle^2},
\end{equation}
where $\langle \Delta \theta^2 \rangle$, $\langle \Delta \phi^2 \rangle$ and $\langle \Delta \theta \Delta \phi \rangle$ are given in Eq.~\ref{eq:sigma_ij}. The luminosity distance uncertainty $\Delta d_L / d_L$ can also be directly derived by Eq.~\ref{eq:sigma_ij}.

Many studies have analyzed the localization accuracy of coalescing massive BBHs for LISA, Taiji or TianQin using the Fisher information matrix approach~\cite{cutler1998angular,moore2002angular,seto2004strong,lang2006measuring,lang2008localizing,trias2008lisa,thorpe2009lisa,Ruan:2019tje,wang2020numerical,zhang2021parameter,shuman2022massive}. 
Their results indicate that it is not enough for a single space-based GW detector to uniquely identify the host galaxy of a non-precession massive BBH from inspiral GW observations.
However, considering that LISA, Taiji and TianQin would launch in the same period, it is feasible to conduct joint GW observations with multiple detectors.
The performance of the multi-detector network on sky localization can also be estimated by above equations.
The results in Ref.~\cite{seto2004strong} indicate that the localization accuracy of massive BBHs can be dramatically improved by a network of two or three LISA-like detectors.
Specifically, some authors present the demonstration of the LISA-Taiji network in Ref.~\cite{Ruan:2020smc}.
For the network, the configuration angle subtended by the heliocentric orbit between LISA and Taiji is set at $40^\circ$ (see Fig.~\ref{fig:LISA_Taiji}).
The LISA constellation follows the Earth by $20^\circ$, while the Taiji constellation moves ahead of the Earth by $20^\circ$.
In fact, the angular resolution of massive BBHs for the LISA–Taiji network depends on the configuration angle.
According to the analysis of 10,000 equal-mass BBHs distributed at different sky positions with total intrinsic mass $M = 10^5 M_\odot$ and redshift $z=1,3$, the angular resolution will reach a minimum value at a configuration angle of $180^\circ$~\cite{Ruan:2020smc}.
As the configuration angle increases from $3^\circ$ to $40^\circ$, the median value of angular resolution is improved by about $2$ orders of magnitude.
Nevertheless, the value is only improved by about $0.6$ orders of magnitude as the configuration angle increases from $40^\circ$ to $180^\circ$.
Thus, considering the potential cost of placing LISA or Taiji in space, it is reasonable to set the configuration angle at $40^\circ$ for accurately localizing the GW sources.
Moreover, for the LISA-Taiji network, the localization accuracy also depends on the inclination of the triangular constellation formed by three spacecraft.
Some calculations suggest that the LISA-Taiji network has better performance on sky localization for massive BBHs when the triangular constellation of Taiji has a $-60^\circ$ inclination compared to the LISA~\cite{Wang:alternative1,shuman2022massive}.

\begin{figure*}[!htbp]
\center
\includegraphics[height=5.3cm]{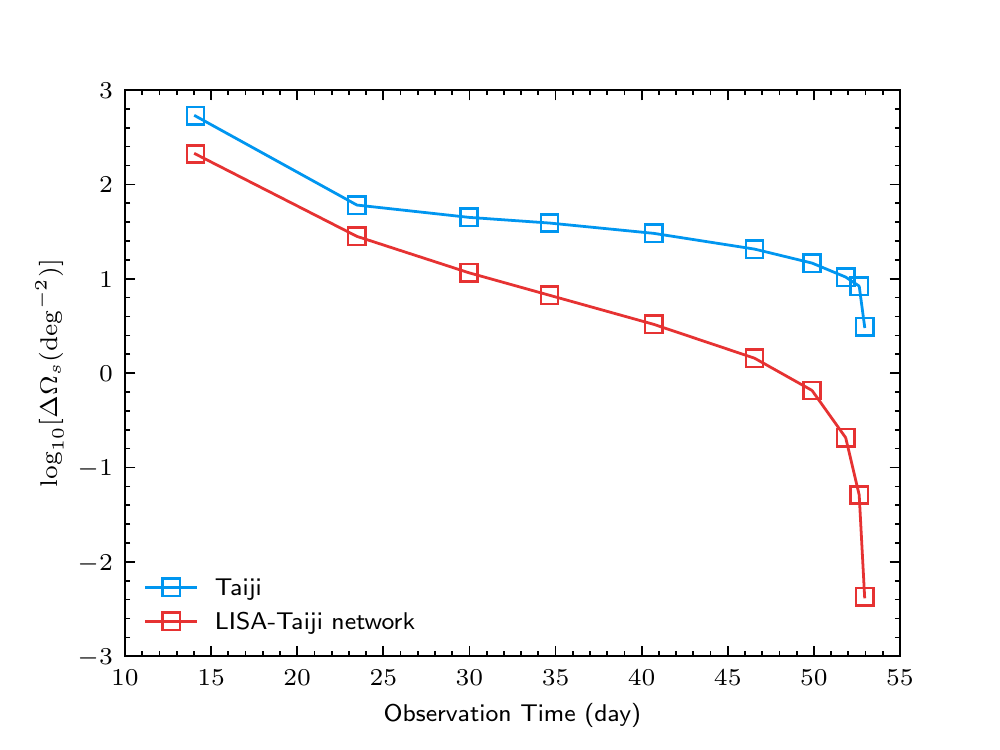}
\includegraphics[height=5.3cm]{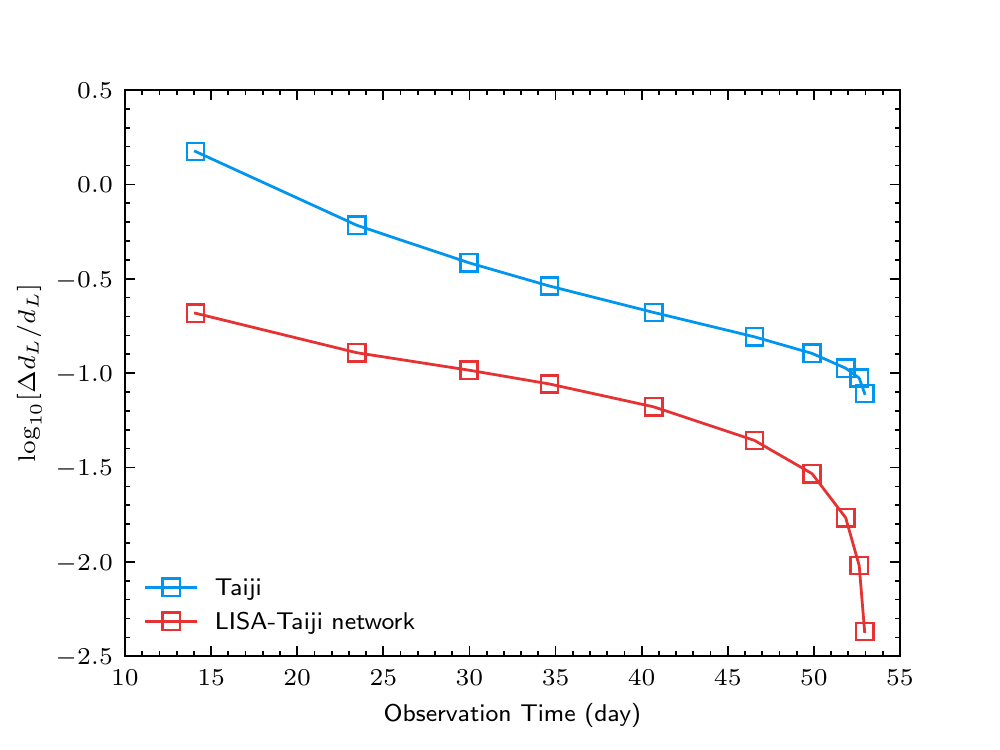}
\caption{The angular resolution $\Delta \Omega_s$ (left panel) and the luminosity distance uncertainty $\Delta d_L / d_L$ (right panel) as a function of the observation time for Taiji (blue) and the LISA-Taiji network (red). The GW source is chosen to an equal-mass BBH with total intrinsic mass $M = 10^5 M_\odot$ and redshift $z=1$. Figures from Ref.~\cite{Ruan:2019tje}.}
\label{fig:local_time}
\end{figure*}
\begin{figure*}[!htbp]
\center
\includegraphics[height=5.3cm]{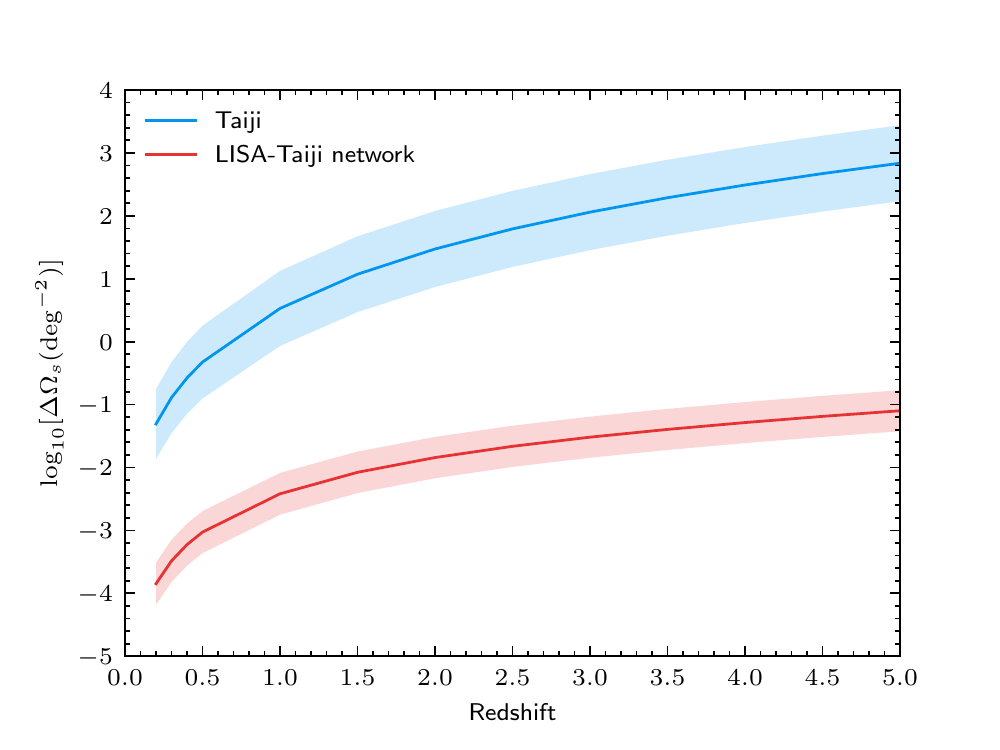}
\includegraphics[height=5.3cm]{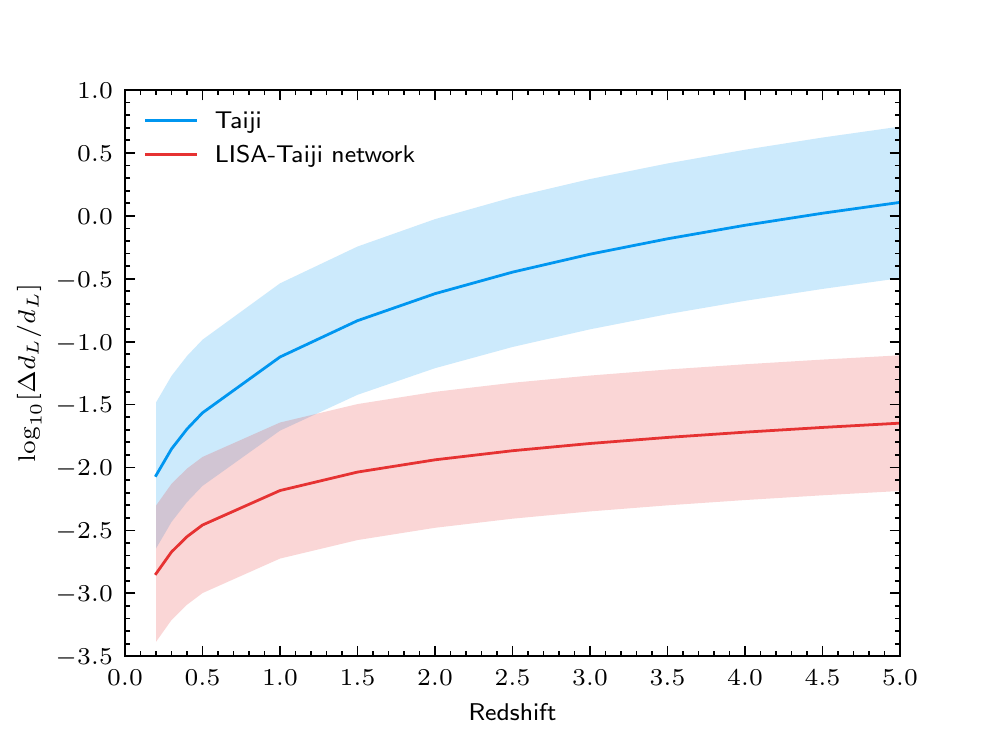}
\caption{The angular resolution $\Delta \Omega_s$ (left panel) and the luminosity distance uncertainty $\Delta d_L / d_L$ (right panel) as a function of redshift for Taiji (blue) and the LISA-Taiji network (red). The center line denotes the median value while the shaded region denotes the $1\sigma$ uncertainty using a catalogue of 10,000 simulated equal-mass BBHs with a fixed redshift at different sky positions and inclination angles. The total intrinsic masses of massive BBHs are fixed at $M = 10^5 M_\odot$. Figures from Ref.~\cite{Ruan:2019tje}.}
\label{fig:local_redshift}
\end{figure*}

The network of multiple space-based detectors provide a prospect to identify the host galaxy of massive BBHs before the final merger.
For a given GW signal observed 52 days prior to merger, Taiji can localize the massive BBH with $\Delta \Omega_S < 4 \ \text{deg}^2$ and $\Delta d_L / d_L < 8\%$ through the observation of the inspiral phase, while the LISA-Taiji network can localize the massive BBH with $\Delta \Omega_S < 0.005 \ \text{deg}^2$ and $\Delta d_L / d_L < 0.5\%$ (see Fig.~\ref{fig:local_time})~\cite{Ruan:2019tje}. 
In this case, the GW source is an equal-mass BBH with total intrinsic mass $M = 10^5 M_\odot$ and redshift $z=1$.
The LISA-Taiji network is capable of narrowing the source localization region by about four orders of magnitude compared to an individual detector, and the conclusion also applies to the cases of $M = 10^6 M_\odot$ and $M = 10^7 M_\odot$~\cite{Ruan:2019tje}.
For massive BBHs with different inclination angles and sky positions, the LISA-Taiji network can still provide great improvement on the localization accuracy.
Fig.~\ref{fig:local_redshift} shows the angular resolution $\Delta \Omega_s$ (left panel) and the luminosity distance uncertainty $\Delta d_L / d_L$ (right panel) as a function of redshift for Taiji and the LISA-Taiji network. 
For a fixed redshift in the figure, the median value and the $1\sigma$ uncertainty are calculated from 10,000 simulated equal-mass BBHs with fixed total intrinsic mass $M = 10^5 M_\odot$ but random inclination angles and sky positions.
Assuming a uniform distribution of galaxies in comoving volume with a number density of $0.02 \ \text{Mpc}^{-3}$, the LISA-Taiji network is capable of identifying the host galaxy of the massive BBH with total intrinsic mass $M = 10^5 M_\odot$ when satisfying the galaxy redshift $z < 0.75$~\cite{Ruan:2019tje}.
It is estimated that the host galaxies of massive BBHs with $M = 10^6 M_\odot$ and $M = 10^7 M_\odot$ can be identified for $z < 0.96$ and $z < 0.45$, respectively.
Furthermore, for unequal-mass BBHs with component masses $(10^7, 3.3\times 10^6)M_{\odot}$, $(10^6, 3.3\times 10^5)M_{\odot}$ and $(10^5, 3.3\times 10^4)M_{\odot}$, the LISA-Taiji network can improve the angular resolution by more than one order of magnitude compared to individual LISA or Taiji~\cite{wang2020numerical}.
In addition to the LISA-Taiji network, the LISA-TianQin network can also bring comparable improvement on localization accuracy of massive BBHs, and it will be further improved by several times with the network of all three detectors~\cite{zhang2021parameter,shuman2022massive,Zhang:2021wwd}.

\subsection{Localization accuracy of Galactic binaries}
Numerous compact binaries in the Milky Way are also potential sources of space-based GW detectors~\cite{LISA:2017pwj, ruan2020taiji,hu2017science}.
The components of these GBs can be white dwarfs, neutron stars and stellar-mass black holes. 
The GBs emit continuous and nearly monochromatic GW signals which are highly overlapped in the sensitive band of space-based GW detectors.
Although most of the signals form an unresolved foreground for the detectors, it is expected that tens of thousands will be individually resolved with source parameters~\cite{korol2017prospects,lamberts2019predicting,liu2023confusion}.
Moreover, some GBs known as ``verification binaries" have already been detected through electromagnetic observations, which provides guaranteed sources for joint gravitational and electromagnetic observations~\cite{kupfer2018lisa}.
The Fisher information matrix approach can also be used to estimate the localization accuracy of the resolved GBs.
Due to the differences of the arm length and the motion of detector, LISA and Taiji has better localization accuracy and better sky coverage for the sources emitting GW signals at $1 \ \text{mHz}$ and $10 \ \text{mHz}$, while TianQin has a better performance on the sources emitting GW signals at $100 \ \text{mHz}$~\cite{jiang2023sky}.
The network of the two or three detectors is capable of improving the precision of sky localization for signals at frequencies in the range of $1-100 \ \text{mHz}$ by a factor of 2 to 10 comparing to a single detector and provides larger sky coverage for the angular resolution~\cite{wang2020numerical,Zhang:2020hyx,jiang2023sky}.

\subsection{Cosmological implication in GW sirens}
\begin{figure}
 \includegraphics[width=8.5cm]{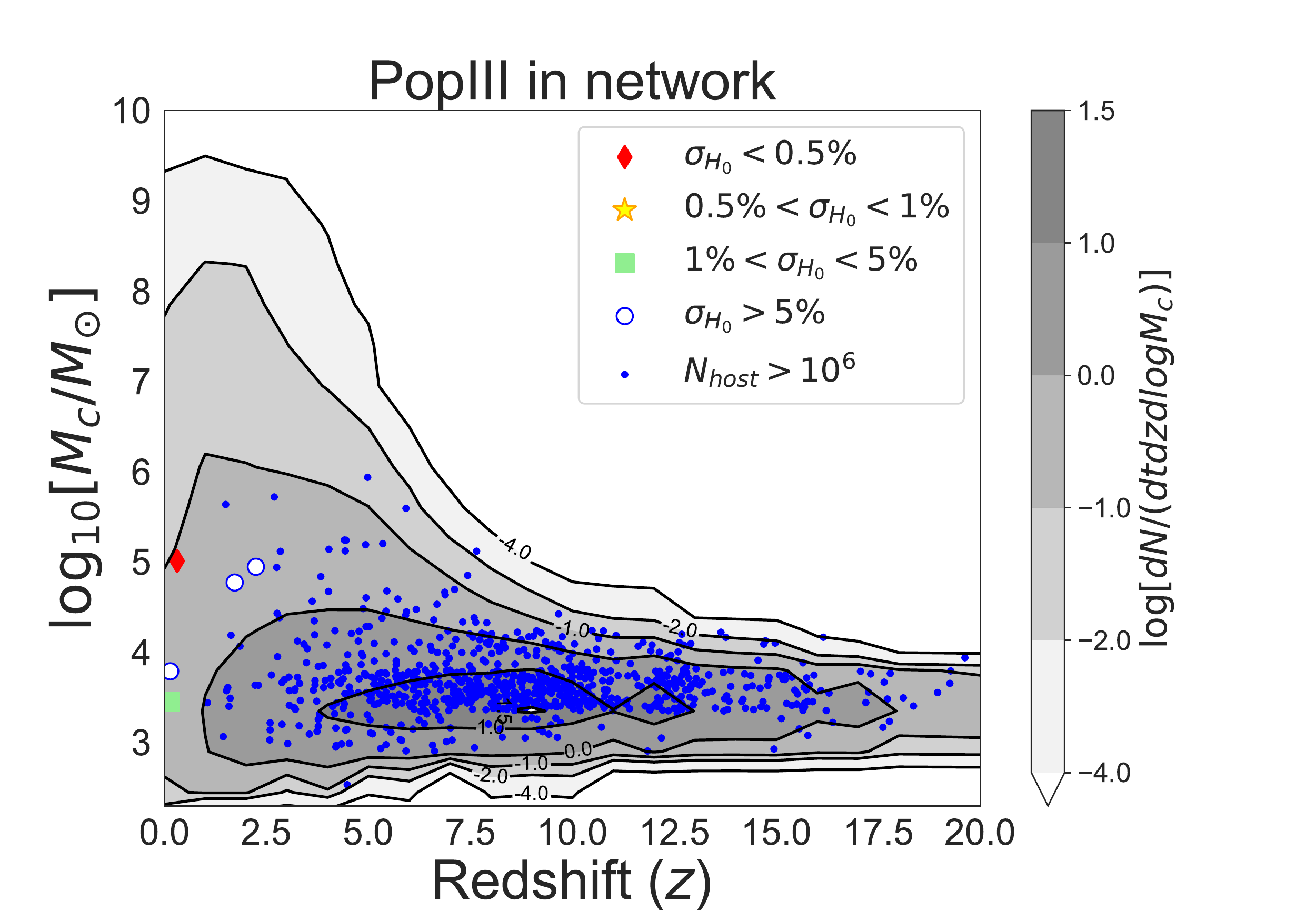}
 \includegraphics[width=8.5cm]{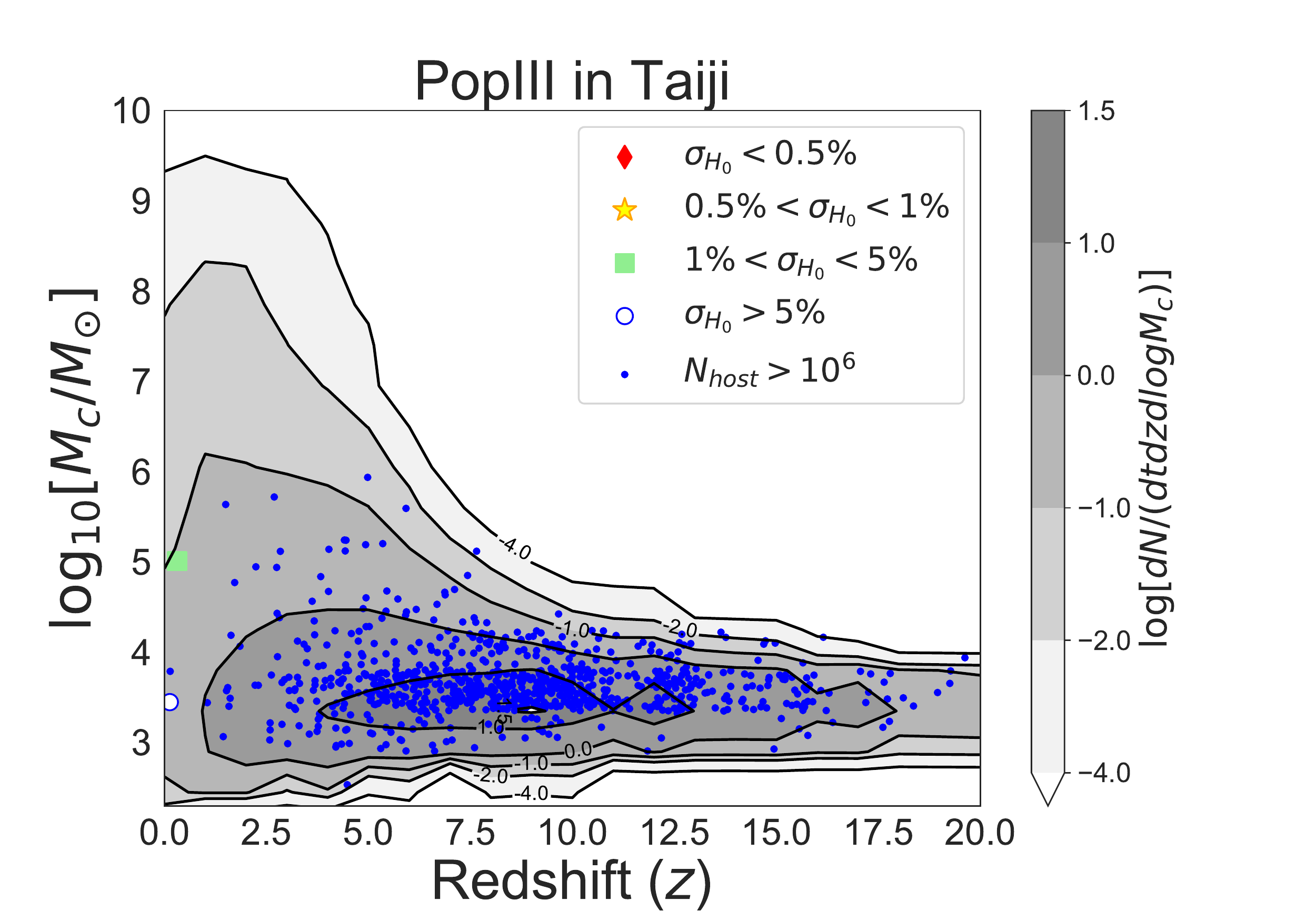}
 \includegraphics[width=8.5cm]{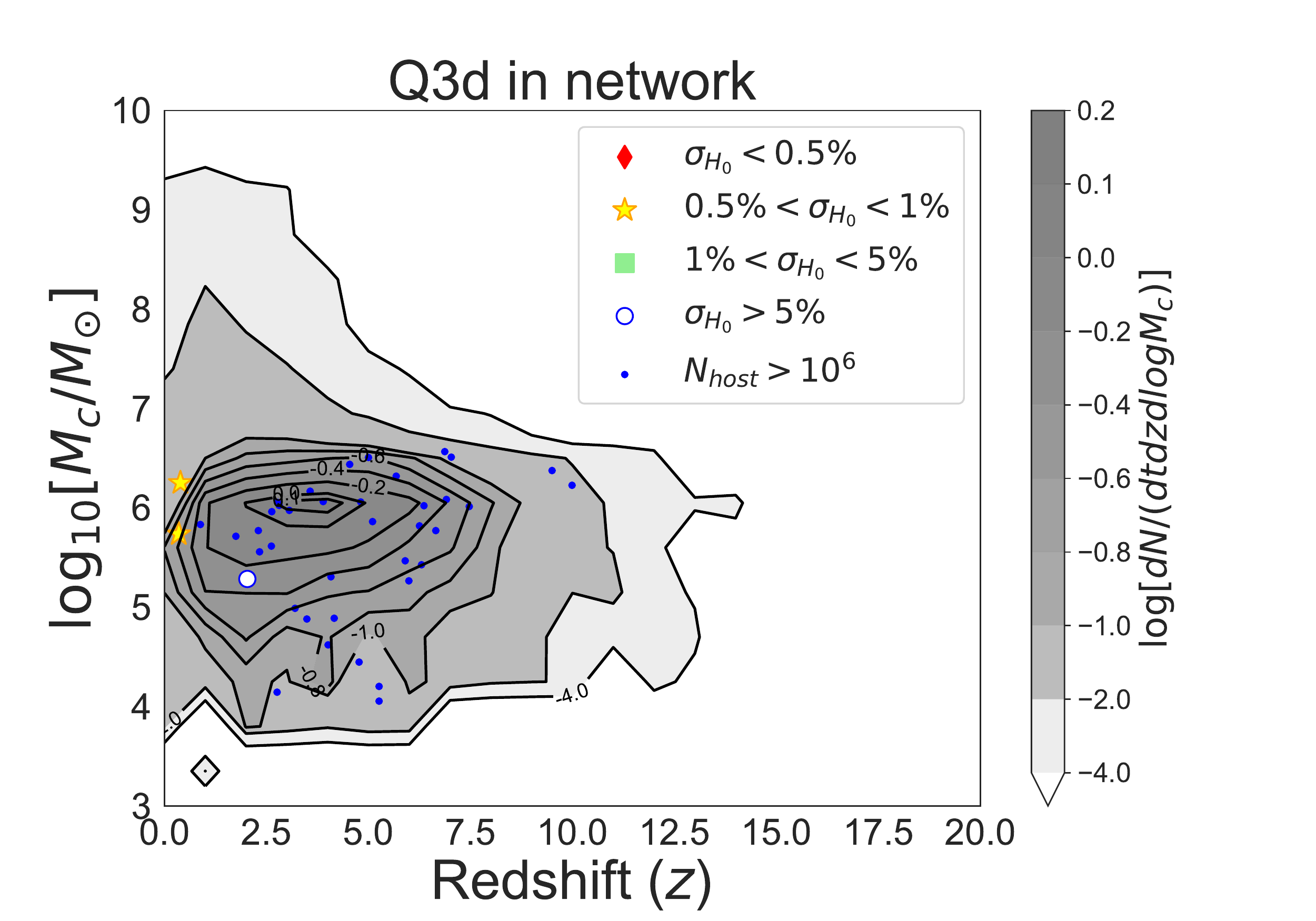}
 \includegraphics[width=8.5cm]{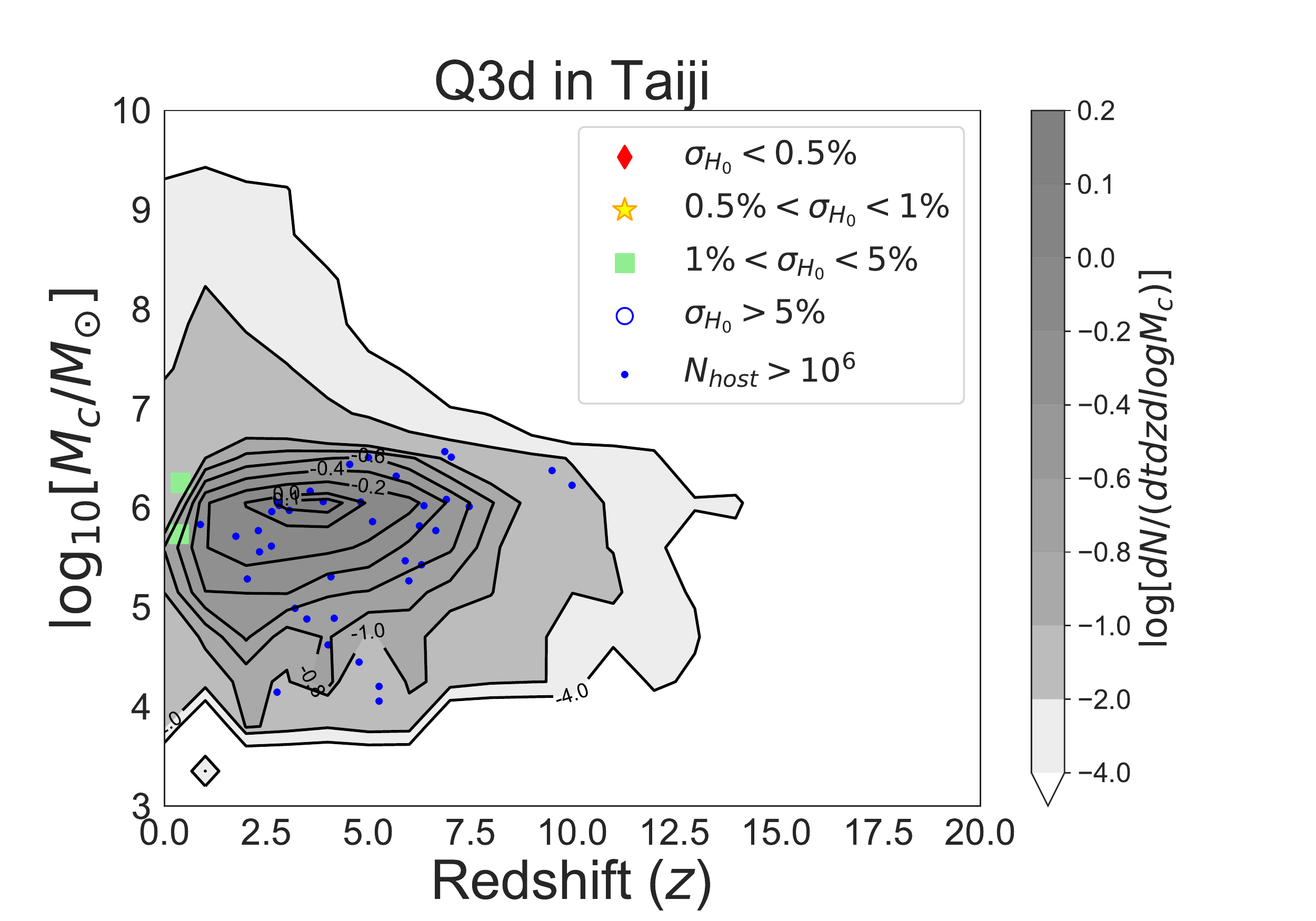}
 \includegraphics[width=8.5cm]{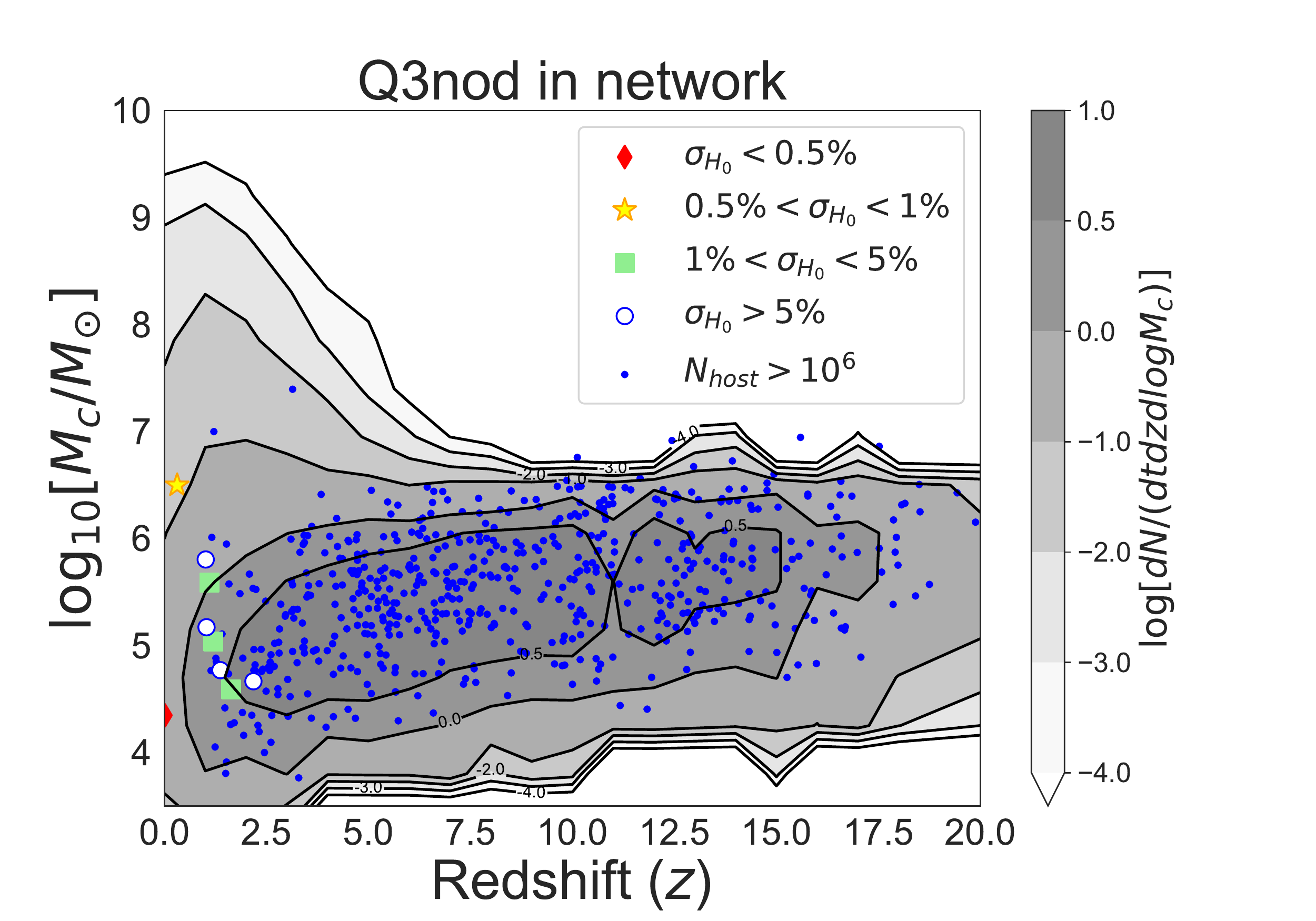}
 \includegraphics[width=8.5cm]{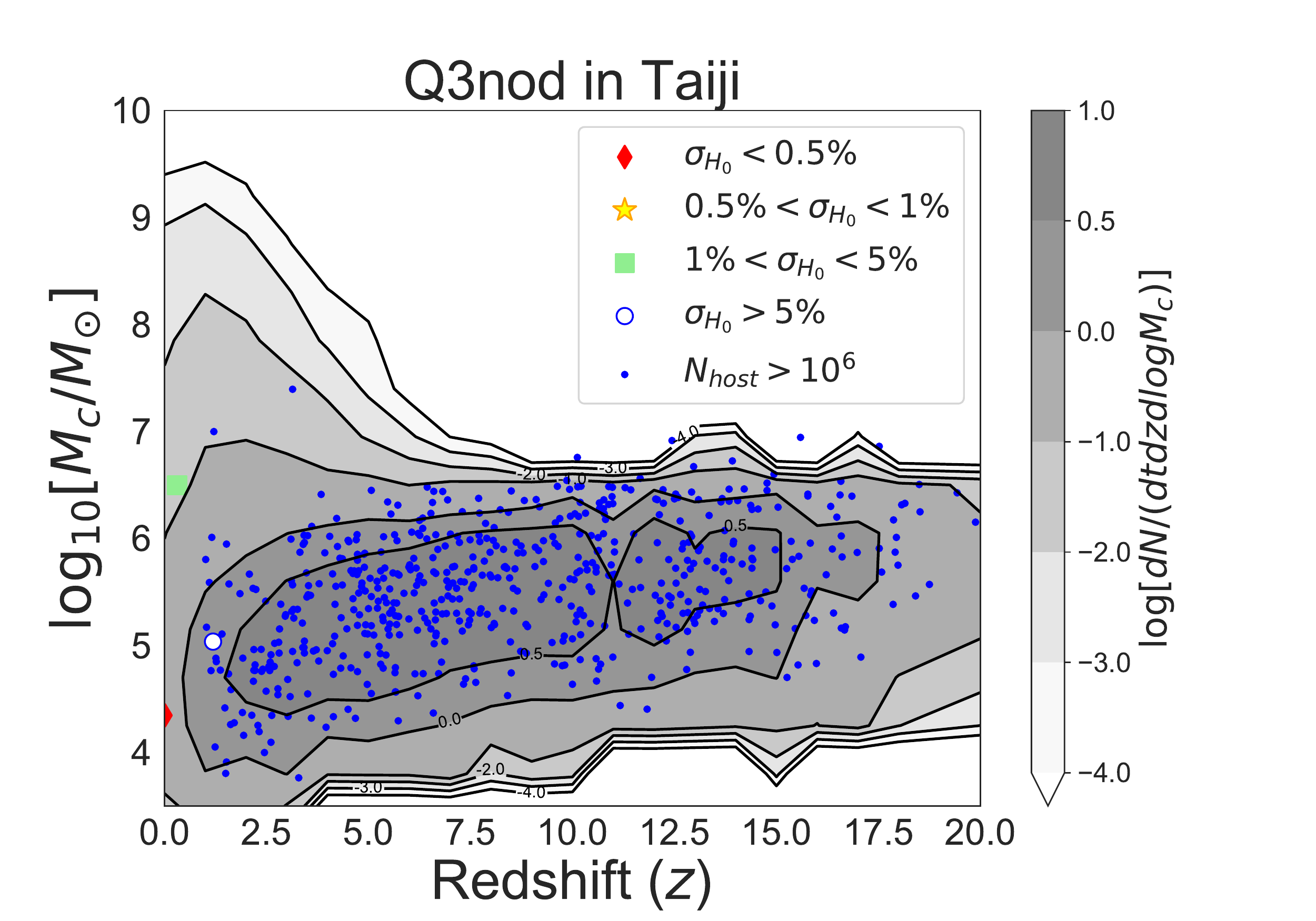}
 \caption{The simulated merger event rate distribution of massive BBHs in redshift and chirp mass within 5-year observation time of the LISA-Taiji network and Taiji-only mission.
Red diamonds ($\sigma_{H_0}/H_0<0.5\%$), yellow stars ($0.5\%-1\%$), green squares ($1\%-5\%$) as well as blue circles ($>5\%$) are the classified dark sirens according to their Hubble parameter
estimation accuracies. The filled blue spots are the unqualified dark sirens whose possible host galaxy numbers are more than $10^6$
due to the poor sky localisation. The background grey contours are the theoretical massive BBH merger event rate distribution.
The first column are the results in the LISA-Taiji network, while the second column are for the  Taiji-only case.
The first, second and third rows are the predictions from $3$ different massive black hole models, namely PopIII, Q3d and Q3nod, respectively. These plots are from Ref.~\cite{Wang:2020dkc}.}
 \label{fig:hu}
\end{figure}

The better distance and sky localization determination in the network\cite{Ruan:2019tje,Zhang:2020hyx,Zhang:2021wwd} will facilitate the cosmological implications, such as the GW sirens\cite{Wang:2020dkc,Wang:2021srv,Yang:2021qge}. The GW siren method for distance measurement can be classified into two categories, namely the bright\cite{Schutz:1986gp,Holz:2005df} and dark\cite{Chen:2017rfc,Fishbach:2018gjp,Gray:2019ksv} sirens. The former relies on the EM counterpart observations to get the redshift information of the host galaxies. Due to the existence of the gas rich accretion disk, we do expect the EM counterpart flare from the massive BBH, for which can be used as the bright siren\cite{Holz:2005df}. Even in the case that the flare is too faint or the jet does not point to us, we can still use the statistical method to infer the redshift of the host galaxy, i.e., the dark siren method. There are various ways to infer the redshift, the ``statistical galaxy catalog method'' is one of the most faithful method\cite{LIGOScientific:2021aug}. The advantage of the dark siren with respect to the bright siren is that the former is much less expensive. It does not ask for the follow-up monitoring at all, but only the existence of the galaxy catalogs in the targeted sky area. The dis-advantage of the dark siren is that the redshift is completely degenerate with the distance without knowing the Hubble parameter. Hence, for single dark siren system, the distribution of the Hubble parameter posterior is multi-peaked. In order to have an accurate estimation of the Hubble parameter, we need to combine many dark siren systems. As shown in Fig.~\ref{fig:hu}, the network of LISA-Taiji (wrt. Taiji-only) can significantly improve the numbers of good GW sirens, whose distance determination errors are less than $5\%$\cite{Wang:2020dkc}. For bright sirens, the LISA-Taiji network could double the detected event numbers wrt. Taiji-only\cite{Wang:2021srv}. As shown by \cite{Tamanini:2016zlh}, in fact the counterparts detectable by LSST are always detectable by SKA+ELT. The number of LSST counterpart detections is around 1. According to the calculation in \cite{Yang:2021qge}, within 5 years observation of massive BBH GW events by LISA, the number of the radio counterparts observed by SKA is 34; the number of the optical observations with ELT of SKA counterparts hosts is 28. For 5 years of observation of the LISA-Taiji network, the number of SKA+ELT counterparts is around 58 (if the average improvement of localization is 100 times the number is 83)\cite{Yang:2021qge}. Besides the Hubble parameter estimation, the network can also tell the black hole formation mechanism\cite{Yang:2022cgm}. In particular, the joint network has the potential to observe growing light seeds in the range $15<z<20$ while a single detector can hardly see, which would shed light on the light seeding mechanism. 

\section{Testing parity violation} \label{sec:test_pv}
In this section, we discuss the correlation analysis for SGWBs with space detector networks, paying special attention on the measurement of their Stokes $V$ parameter.  As in the case of electromagnetic waves, the Stokes $V$ parameter is defined as the asymmetry between the right- and left-handed circular polarization modes and is closely related to parity violation processes. 
 
Currently, these are many theoretical scenarios for generating chiral asymmetries ($ V\ne0$) for SGWBs, including those with the Chern-Simons term (see e.g., \cite{Alexander:2004us,Satoh:2007gn,Obata:2014loa,Adshead:2012kp,Kahniashvili:2005qi,Ellis:2020uid}). With its high transparence, a SGWB could be an important fissile from the early universe, and an observational confirmation of its chiral asymmetry would have a considerable impacts on cosmological studies. Considering the observed isotropy of our universe, we set the monopole components of SGWBs as our primary target in the following.

\subsection{monopole pattern}
We  decompose the metric perturbation induced by SGWB in terms of Fourier modes as 
\begin{equation}
\veh(t,\vex)=\sum_{P=R,L} \int^{\infty}_{-\infty} df \int_{S^2} d\ven~
h_P(f\ven)  \veb^P_{}(\ven) e^{2\pi i f (\ven \cdot \vex-t) }.
\end{equation}
Here    the unit vector $\ven$  represents the wave propagation direction and the two tensors  $\veb^{R,L}(\ven)$
show  the right- and left-handed circular polarization bases.  They are given by  the linear polarization bases $\veb^{+,\times}$ (the plus and cross modes) as 
\begin{equation}
\veb^R=\frac{(\veb^++i \veb^\times)}{\sqrt{2}},~~~\veb^L=\frac{(\veb^+-i \veb^\times)}{\sqrt{2}} \label{+x}.
\end{equation}  

As mentioned earlier, our main target here is the isotopic components of the background.  We have two relevant modes, corresponding to the Stokes $I$ and $V$ parameters. They are related to the mode spectra as
\begin{eqnarray}
\left(
  \begin{array}{c}
     \lla h_R(f\ven) h_R(f'\ven')^* \rra +\lla h_L(f\ven) h_L(f'\ven')^* \rra \\
     \lla h_R(f\ven) h_R(f'\ven')^* \rra -\lla h_L(f\ven) h_L(f'\ven')^* \rra 
  \end{array}
 \right)
=\frac{ \delta_{{\ven,\ven'}}\delta_{f,f'} }{4\pi}
\left(
  \begin{array}{c}
     I(f)\\
     V(f) 
  \end{array}
 \right) \label{rl}
\end{eqnarray}
with the Delta functions \cite{Seto:2006hf}. In Eq. (\ref{rl}), the parameter $I$ is given as the sum of the right- and left-handed waves, representing the total intensity.   On the other hand, the parameter $V$ characterizes their asymmetry.  
 The other two Stokes parameters $Q$ and $U$ do not have monopole modes, as they are related to linear polarization (introducing specific spatial direction). 
Using the normalized energy density $\Omega_{\rm GW}(f)$  and the prioritization degree  $\Pi(f)$, we can put 
\begin{equation}
I(f)=\frac{\rho_c}{4\pi^2f^3} \Omega_{\rm GW}(f), ~~V(f)=\frac{\rho_c}{4\pi^2f^3} \Omega_{\rm GW}(f)\Pi(f)
\end{equation}
with the critical density of the universe $\rho_c$.

\subsection{correlation analysis}

Next, let us consider a detector $a$ (at the position $\vex_a$ with the detector tensor ${\bf d}_a$)  under the  low frequency approximation. 
In the Fourier space, its reaction to the SGWB can be expressed as 
\begin{equation}
h_a(f)=\sum_{P=R,L} \int d\ven \,h_P(f\ven)({\bf d}_a:\veb^{P}) e^{2\pi i f \ven \cdot \vex_a }
\end{equation}
 with the colon $:$ for the double contraction of the two involved tensors.
Then we consider another noise independent interferometer $b$ and take a correlation product  $\lla h_a(f) h_b(f\rq{})^*\rra =C_{ab}(f)\delta_{f,f\rq{}}$.  Using Eqs. (\ref{+x}) and (\ref{rl}), we have
\begin{equation}
C_{ab}(f)=\frac{8\pi}5 \lkk \gamma_{Iab}(f)I(f)+\gamma_{Vab}(f) V(f) \rkk. 
\end{equation}
Here $\gamma_{Iab}$ and $\gamma_{Vab}$ are called the overlap reduction functions, respectively showing the sensitivity to the $I$ and $V$ modes. After some algebra, we have
\begin{equation}
\gamma_{Iab}(f)=\frac5{8\pi} \int d\ven (F_a^+ F_b^++F_a^\times F_b^\times) e^{2\pi i f \ven\cdot (\vex_a-\vex_b)},~~\gamma_{Vab}(f)=\frac{-5i}{8\pi} \int d\ven(F_a^+ F_b^\times -F_a^\times F_b^+)e^{2\pi i f \ven\cdot (\vex_a-\vex_b)}
\end{equation}
with the standard beam patter functions ($F_a^{+,\times}(\ven)\equiv {\bf d}_a:\veb^{+,\times}(\ven)$) as  \cite{Flanagan:1993ix,Allen:1997ad,Seto:2006hf}. We discuss the these overlap reduction  functions for the LISA-Taiji and LISA-TianQin networks in the following two subsections. 

\subsection{LISA-Taiji network}

The LISA-Taiji network has an interesting geometrical symmetry \cite{Seto:2020zxw}.  In addition, a triangular detector like LISA has an  internal symmetry at composing orthogonal data sets.    By suitably combining these two symmetries, we can use the LISA-Taiji network as an attractive tool to explore the parity properties of a SGWB (see also \cite{Orlando:2020oko,Omiya:2020fvw,Liu:2022umx}).

\begin{figure}[htb]
\centering
 \includegraphics[width=0.8\textwidth]{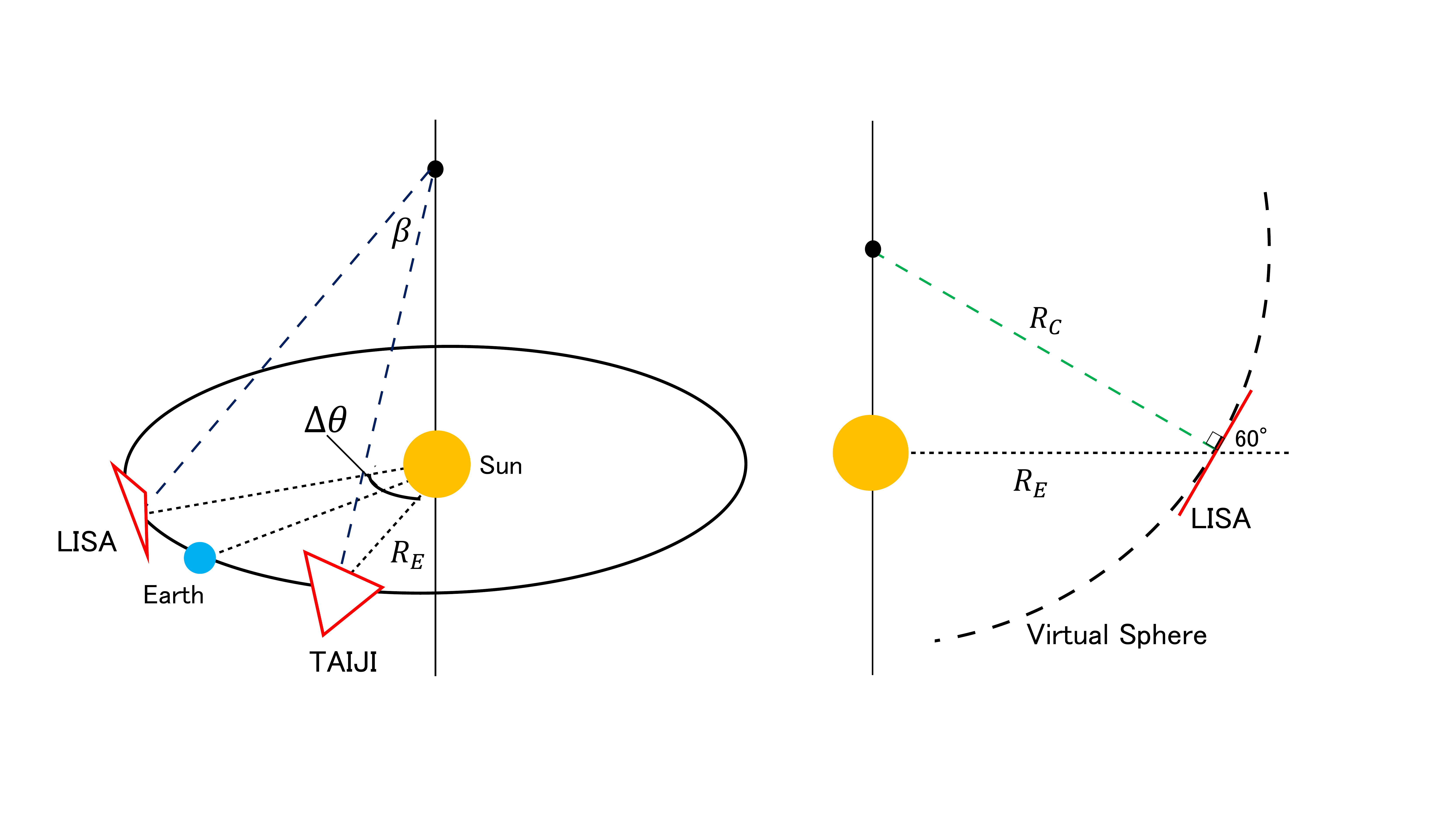}
 \caption{The global geometry of the LISA-Taiji network. The two detector planes are tangential to a fixed virtual sphere of radius $R_C=2 R_E/\sqrt{3}$ ($R_E=1$a.u.: the Sun-Earth distance).  The orbital phase angle is $\Delta  \theta=40^\circ$, corresponding t o the opening angle $\beta=34.46^\circ$ measured from the center of the virtual sphere.    }
 \label{fig:s1}
\end{figure}

\begin{figure}[htb]
\centering
 \includegraphics[width=0.8\textwidth]{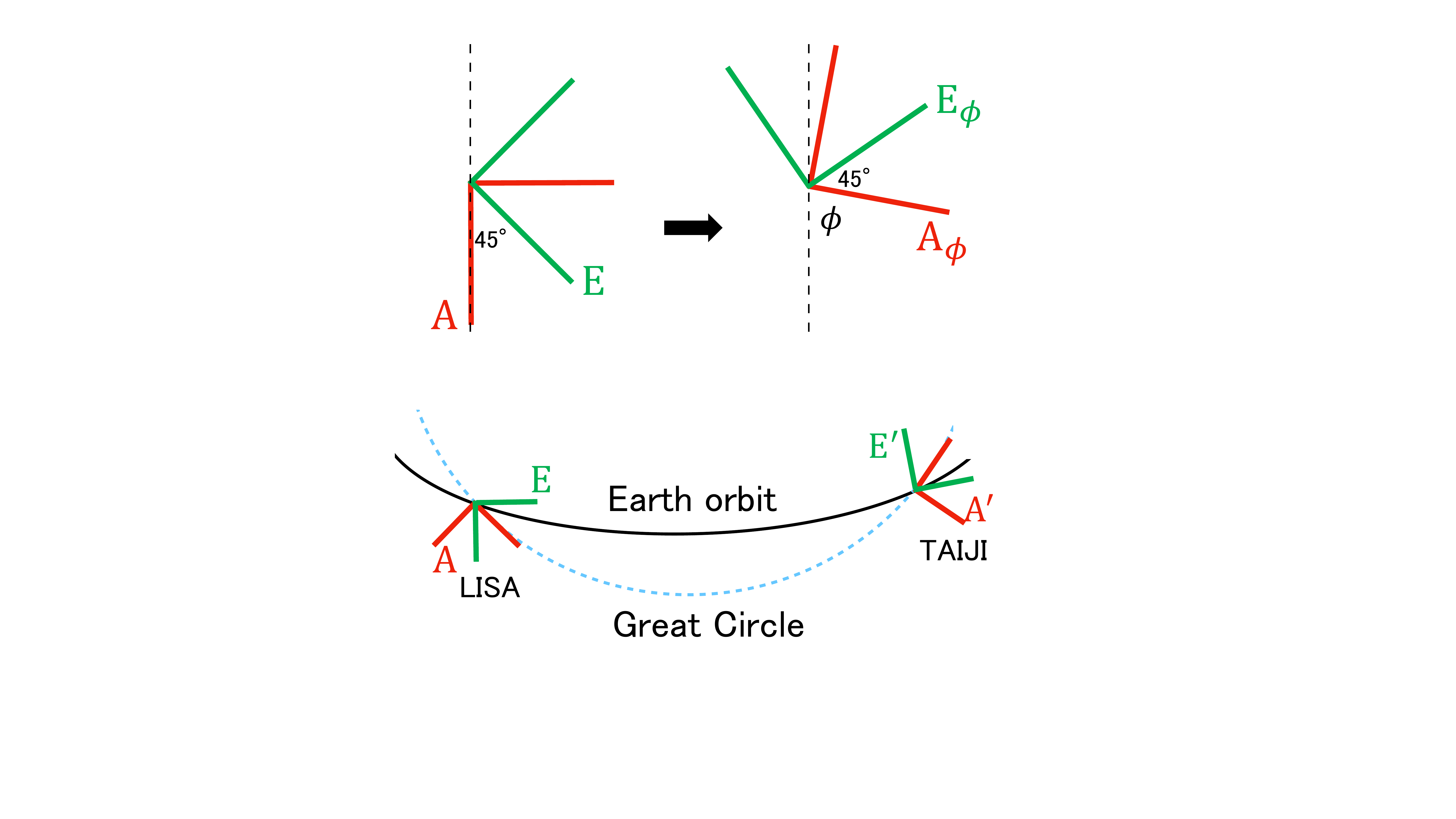}
 \caption{ (Top) Orientation of the two effective L-shaped
interferometers A and E on the detector plane. By taking their linear combination Eq.(\ref{comb}), we can virtually rotate them with the new data channels $(A_\phi,E_\phi)$. (Bottom) By using the freedom of the rotational angle $\phi$, we can align the interferometer A with respect to the great circle on the virtual sphere.
}
 \label{fig:s2}
\end{figure}

As shown in Fig.~\ref{fig:s1}, the detector plane of LISA is tangential to a virtual sphere of radius $R_C=2/\sqrt3 R_E=1.15$\,a.u. ($R_E=1$\,a.u.: the Sun-Earth distance).  We have two options for the orientation of Taiji's detector plane. If we take the standard choice, Taiji is also tangential to LISA's virtual sphere (see Fig.~\ref{fig:s1} and also \cite{Seto:2020mfd,Wang:alternative2}) From the center of the sphere,  the two detectors are separated by the angle $\beta=2\sin^{-1}[\sqrt{3}\sin(\Delta \theta/2)/2]))=34.5^\circ$ for the orbital phase difference $\Delta\theta=40^\circ$.  Since a sphere is highly symmetric, it will greatly help us  to figure out the underlying geometrical structure of the network. 

Here we mention the additional symmetry of the system. From a LISA-like triangular detector unit, we can compose three noise orthogonal data $A, E$ and $T$ \cite{Prince:2002hp}. The T mode is much less sensitive in the frequency regime relevant for our study.  Due to a systematic cancellation, we cannot probe the  monopole components of a SGWB by correlating these modes.  
As shown in Fig.~\ref{fig:s2}, in the low frequency regime, the $A$ and $E$ modes correspond to a pair of two L-shaped interferometer with the offset angle $45^\circ$, composing a symmetric base system.  Importantly, by taking their linear combinations, 
\begin{equation}
A_\phi=A\cos 2\phi+E\sin 2\phi,~~E_\phi=-A\sin 2\phi+E\cos 2\phi, \label{comb}
\end{equation} 
we can effectively rotate the two interferometers by the angle $\phi$ (see  Fig.~\ref{fig:s2}). 

Now we couple this internal symmetry with the virtual sphere of the LISA-Taiji network.  We can align the orientation of the effective L-shaped interferometers by using the great circle (geodesic) that connects LISA and Taiji on the sphere (see Fig.~\ref{fig:s2}). We put the resultant effective interferometers simply by $A$ and $E$ for LISA,  and $A'$ and $E'$ for Taiji.  Then, because of the mirror symmetry  of the pairwise interferometers with respect to the plane containing the great circle (and the center of the sphere), we can readily find that the combinations $AA'$ and $EE'$ are sensitive only to the parity even components (i.e $\gamma_{VAA'}=\gamma_{VEE'}=0$). We similarly find  $\gamma_{IAE'}=\gamma_{IAE'}=0$ and  $\gamma_{VEA'}=\gamma_{VAE'}$.  Furthermore, we can directly use the analytical expressions for the overlap reduction functions derived originally for the ground based detectors \cite{Flanagan:1993ix,Allen:1997ad,Seto:2007tn}. For example, we have
\begin{equation}
\gamma_{VEA'}(f)=\gamma_{VAE'}(f)=\sin\frac{\beta}2 \lkk \lmk   -j_1+\frac78 j_3\rmk+ \lmk   j_1+\frac38 j_3\rmk\cos\beta   \rkk
\end{equation}
with the spherical Bessel function $j_l(y)$ and its argument $y=\pi f c^{-1} R_C\sin(\Delta \theta/2)$.

\begin{figure}[htb]
\centering
 \includegraphics[width=0.6\textwidth]{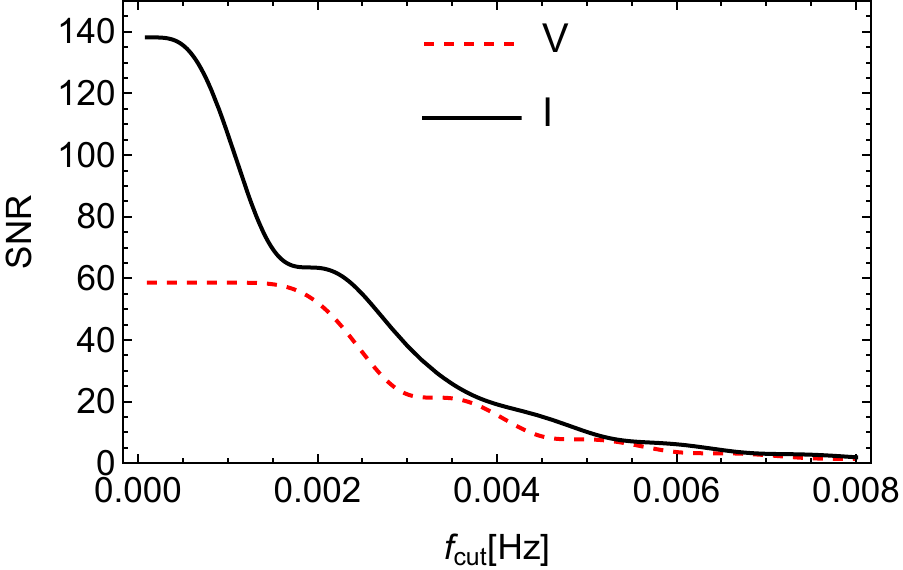}
 \caption{Signal-to-noise ratios for the $I$ and $V$ modes with $f_{\rm max}=10$\,mHz and $T_{\rm obs}= 10$\,yr.  We set $\Omega_{\rm GW}(f)=10^{-11}$ and  $\Pi(f)\Omega_{\rm GW}(f)=10^{-11}$ for the background spectra. }
 \label{fig:s3}
\end{figure}

We can now evaluate the signal-to-noise ratios $SNR_{I,V}$  with the LISA-Taiji network, using standard expressions for the correlation analysis 
\begin{eqnarray}
SNR_I^2=\lmk \frac{16\pi}5  \rmk^2 T_{\rm obs}  \lkk 2\int_{f_{\rm min}}^{f_{\rm max}}  df \frac{(\gamma_{IAA}^2+\gamma_{IEE}^2) I(f)^2}{f^6S_{\rm L}(f)S_{\rm T}(f)}   \rkk,   \\
SNR_V^2=\lmk \frac{16\pi}5  \rmk^2 T_{\rm obs}  \lkk 2\int_{f_{\rm min}}^{f_{\rm max}}  df \frac{ 2\gamma_{VAE}^2  V(f)^2}{f^6S_{\rm L}(f)S_{\rm T}(f)}   \rkk  .
\label{snr}
\end{eqnarray}
Here $S_{\rm L}(f)$ and $S_{\rm T}(f)$ represent the noise spectra of LISA and Taiji, and we are assumed to use the frequency regime  $[f_{\rm min},f_{\rm max}]$. In Fig.~\ref{fig:s3}, we show our numerical results for flat spectral models  $\Omega_{\rm GW}=10^{-11}$ and $ \Pi \Omega_{\rm GW}=10^{-11}$.   The step-like patterns are mainly caused by the frequency dependence  of the overlap reduction functions, indicating that the importance of the contribution around 2-3mHz.  If we require the minimum  signal-to noise ratio 10,   the detection threshold is $\Omega_{\rm GW}\sim 7\times 10^{-13}$ and  $\Omega_{\rm GW} \Pi \sim 1.6\times 10^{-12}$ for a ten year correlation analysis.

\subsection{LISA-TianQin network}
Unlike the LISA-Taiji network (with the overlapped contact sphere),   the LISA-TianQin network \cite{Seto:2020mfd,Liang:2021bde} does not have a global geometrical symmetry.  While we can still use the internal symmetry of a triangle detector to simplify some of analytical expressions \cite{Seto:2020mfd}, the  parity decomposition with the great circle alignment is now inapplicable.   However, since the relative configuration of LISA and TianQin changes with time, we can effectively obtain the correlation products for many network geometry.  By taking their appropriate linear combinations, we can algebraically  separate the two parity components $I$ and $V$.  

If we want to decompose not only the two tensor components but also relevant  vector and scalar components \cite{Omiya:2020fvw,Omiya:2021zif}, this network might be advantageous, in comparison of the number of the independent correlation products and that of the target polarization components. 

%

\section{Substraction of the galactic foreground \label{sec:sgf}}
According to current models and observations, our Milky Way contains a large number of ultracompact binaries with orbital periods of less than a few hours~\cite{Bender:1997hs,Hiscock:2000jn,Nelemans:2001hp,Lamberts:2019nyk}.
The GWs from more than 20 million GBs will enter the space-based GW observation band simultaneously~\cite{Yu:2010fq,Nelemans:2013yg,Korol:2017qcx,Korol:2020lpq}.
Those binaries are in the inspiral phase, millions of years before merger~\cite{Nelemans:2001hp,Marsh:2011yj}.
It means that the continuously emitted GWs will overlap, resulting in a foreground signal for space-based detections~\cite{Bender:1997hs,Hiscock:2000jn,Seto:2004ji,Barack:2004wc,Timpano:2005gm,Nissanke:2012eh,Cornish:2017vip,Karnesis:2021tsh,Boileau:2021sni,Korol:2021pun,Liu:2023qap}.
Only a few tens of thousands of them are bright enough for data analysis to be subtracted with LISA/Taiji~\cite{Umstatter:2005su,Umstatter:2005jd,Wickham:2006af,Crowder:2006eu,Blaut:2009si,Littenberg:2011zg,Cornish:2017vip,Korol:2017qcx,Karnesis:2021tsh,Korol:2021pun,Lu:2022ywf,Liu:2023qap}.
The rest are unresolvable, forming an effective noise, known as `foreground confusion' or `confusion noise'~\cite{Karnesis:2021tsh}.
The foreground confusion provides the dominant part of the noise in the frequency band from 0.5 to 3 mHz~\cite{Cornish:2017vip,Karnesis:2021tsh,Liu:2023qap}, which will affect the detection of other GW sources~\cite{Robson:2017ayy}.
One of the main science objectives of LISA/Taiji is to study the evolution and distribution of the GBs in our Milky Way using the foreground GW signals~\cite{Marsh:2011yj,Nissanke:2012eh,Adams:2012qw,Nelemans:2013yg,LISA:2017pwj,Breivik:2019oar,LUO2020102918,Georgousi:2022uyt}.
Known from electromagnetic observations, about 40 verification binaries are expected to enter the mHz band, which can be used to evaluate the performance of the mission~\cite{Stroeer:2006rx,Shah:2014nea,Korol:2017qcx,Burdge:2019hgl,Burdge:2019ron,Brown:2020uvh,Kupfer:2023nqx}.
The GW detection in a network will aid in the subtraction of GB signals and suppression of foreground confusion~\cite{Zhang:2022wcp}.

Many studies have been conducted on the foreground GW signals from GBs in the mHz band since LISA was proposed~\cite{Hiscock:2000jn,Nelemans:2001hp,Lamberts:2019nyk,Seto:2004ji,Barack:2004wc,Timpano:2005gm,Cornish:2007if,Yu:2010fq,Nissanke:2012eh,Nelemans:2013yg,Cornish:2017vip,Breivik:2019oar,Boileau:2021sni,Korol:2021pun}.
In Ref.~\cite{Cornish:2017vip}, the authors investigated the GB signals with the new LISA design updated in 2016.
The catalog of the GBs they used contains about 26 million sources in our Milky Way~\cite{Toonen:2012jj}.
For different observation times $T_{obs}=$ 6~months, 1~year, 2~years, and 4~years, the iterative procedure proposed in Ref.~\cite{Timpano:2005gm} was used to subtract the bright sources and estimate the residual confusion noise.
For a 4-year observation, about 29 thousand GBs are marked as `resolvable', whose signal-to-noise ratio $>7$.
The analytic fit for the residual confusion noise was obtained, see Eq.~(3) in Ref.~\cite{Cornish:2017vip}.
Most GBs are resolvable at the frequency of 2 to 3~mHz.
The procedure is carried out under the assumption that the resolvable sources can be perfectly subtracted without residuals.
However, in real data analysis, inaccurate removal will introduce errors in parameter estimation~\cite{Robson:2017ayy}.

The updated GB catalog (contains about 29.8 million sources)~\cite{Korol:2020lpq} was used in Ref.~\cite{Karnesis:2021tsh} to develop the methodology for estimating the confusion noise.
For the observation time $T_{obs}=$ 6~months, 1~year, 2~years, 4~years, and 6~years, the authors used a new analytic form to fit the foreground confusion with the signal-to-noise ratio threshold $=5$ and $7$:
\begin{equation}
  S_{c}(f) = \frac{A}{2}  f^{-7/3} e^{-(f/f_1)^\alpha} \left( 1 + \tanh\left( \left( f_{\rm knee} - f \right)/f_2 \right) \right) .
\end{equation}
The fitting parameters $A$, $\alpha$, $f_1$, $f_{\rm knee}$, and $f_2$ can be found in Table II in Ref.~\cite{Karnesis:2021tsh}.
They found that different methods for smoothing the power spectral density yielded different results.
For a 4-year observation and the signal-to-noise ratio threshold $=7$, the number of resolvable sources with LISA is 22951 or 25025, depending on the smoothing method used.

The authors of Ref.~\cite{Korol:2021pun} revisit this problem by employing an observationally driven population of GBs that differs from the previous population based on binary population synthesis.
The observationally driven population yields about 26 million GB sources in LISA band with 60000 of them being resolvable and a significantly different shape of the foreground confusion.
The number of resolvable sources is $2-5$ times that of the binary population synthesis model.

In Ref.~\cite{Digman:2022jmp}, the iterative procedure has been expanded to account for the effect of the detector orbiting the Sun, which causes the foreground confusion noise to vary throughout the year.

The mission design of Taiji is similar to LISA's with slightly better sensitivity, allowing Taiji to extract foreground signals more effectively~\cite{LISA:2017pwj,Hu:2017mde}.
The authors of Ref.~\cite{Liu:2023qap} investigated the foreground signals and confusion noise for Taiji by using the same catalog~\cite{Korol:2020lpq} of 29.8 million GBs as Ref.~\cite{Karnesis:2021tsh}.
The foreground confusion for Taiji is estimated and fitted in the logarithmic scale by a polynomial function with 6-month, 1-year, 2-year, and 4-year observation times:
\begin{equation}
  S_c(f)=\exp\left(\sum_{i=0}^{5} a_i \left(\log\left(\frac{f}{\rm mHz}\right)\right)^i \right) \, {\rm Hz}^{-1}\, .
\end{equation}
The fitting parameters $a_i$ can be found in Table I in Ref.~\cite{Liu:2023qap}.
Including the foreground confusion noise, the full sensitivity curve of Taiji is slightly lower than LISA's at frequencies of $\leq 0.8$ mHz and around 2~mHz. 
With Taiji for a 4-year observation time, 29633 sources are resolvable (signal-to-noise ratio $>7$).
All GBs resolvable with LISA can also be resolved with Taiji.
And Taiji can subtract $\sim 20 \%$ more GB sources whose distribution in our Milky Way is consistent with that of the resolvable sources with LISA.

Subtraction of the galactic foreground will also benefit from network observations.
The authors of Ref.~\cite{Zhang:2022wcp} investigated resolving GB signals with the network of LISA and Taiji-mod.
The `Taiji-mod' is a simplified concept of Taiji, in which they put another LISA in the place of Taiji to simulate the real Taiji.
So, the arm length and noise model of Taiji-mod are the same as LISA's.
The iterative subtraction method they used is based on the particle swarm optimization algorithm.
They found that the LISA-Taiji-mod network can subtract $\sim 75 \%$ more sources than a single LISA.
By using network observation, the residual confusion noise can be reduced below the level with a single detector.
This implies that network detection has a promising future in galactic foreground subtraction.
Besides, the Fisher matrix analysis for parameter estimation may be insufficient in the multi-source GB resolution problem for the extrinsic parameters.

The space-based detectors will have the potential to detect circumbinary exoplanets orbiting double white dwarfs.
This topic is discussed in Refs.~\cite{Seto:2008di,Wong:2018amf,Tamanini:2018cqb,Danielski:2019rvt} for LISA and Ref.~\cite{Kang:2021bmp} for LISA and Taiji.

Recently, Ref.~\cite{Chen:2021sco} reported the possibility of detecting stellar-mass black hole binaries with space-based detectors and their network.
The results show that the LISA-Taiji network can detect about 10 stellar-mass black hole binaries, demonstrating the possibility of multiband GW observations in the future.

\section{Improvement on the number of detectable BBHs \label{sec:sbbh}}
The merging processes of stellar BBHs include the inspiralling stage at early time with GW radiation in the low-frequency band ($\sim 10^{-4}-1$\,Hz) and the final merger with GW radiation in the high-frequency band ($\sim 10-1000$\,Hz). The ground-based GW observatories LIGO/Virgo have detected more than $90$ mergers of stellar BBHs. A number of such BBHs, inspiralling in the low-frequency band before their final merger, are expected to be detected by space borne GW detectors, such as LISA, Taiji, and TianQin, which will enable multiband observations of GWs \cite{2016PhRvL.116w1102S, 2020PhRvD.101j3027L, 2022RAA....22a5020C}. The low-frequency observations provide important information in advance on when and where a BBH merger, if detectable, will occur for high-frequency GW observations and multiwavelength electromagnetic observations. The multi-band GW observations of stellar BBHs are also extremely helpful for understanding the formation mechanism of stellar BBHs,  cosmic evolution, and  precise test of gravity theory, etc. To realize the potential of multi-band GW observations in practice, it is crucial to have a sufficient number of stellar BBHs being detected by the space borne low-frequency GW detectors, such as LISA, Taiji, TianQin, as well as the network composed of these detectors. Below we mainly consider LISA, Taiji, and the LISA-Taiji network for demonstration purpose.

To estimate how many stellar BBHs  can be detected by the low-frequency GW detectors, we generate mock stellar BBH samples according to the cosmic evolution of the BBH merger rate similar as that in \cite{2021MNRAS.500.1421Z, 2022RAA....22a5020C}. We adopt the merger rate density of stellar BBHs given by the current LIGO/Virgo observations, i.e., $\frac{d^2N}{dV dt}(z=0) (1+z)^{\kappa}$ with the local merger rate density of $\frac{d^2N}{dV dt}(z=0)  =19.1^{+8.6}_{-8.5} \gpcyr$ and redshift evolution slope of $\kappa=2.7$ \cite{2021arXiv211103606T}. We adopt the fiducial power law plus peak model for primary mass and mass ratio distribution given in \cite{2021arXiv211103634T} (see the Power Law + Peak (PP) model therein). We generate $100$ realizations for stellar BBHs and then estimate the SNR for each mock stellar BBH in each realization by the observations of LISA, Taiji, or the LISA-Taiji network over a period of either $4$ or $10$ years. We define a stellar BBH as a ``detectable'' one if its SNR is $\geq 8$ regarding to either LISA, Taiji, or LISA-Taiji.  For different realizations, the expected number of ``detectable'' stellar BBHs are different. We rank the $100$ realizations according to the expected number of ``detectable'' stellar BBHs from small to large, and take that with $50$th rank as the representative one.

\begin{figure}
\centering
 \includegraphics[width=0.7\textwidth]{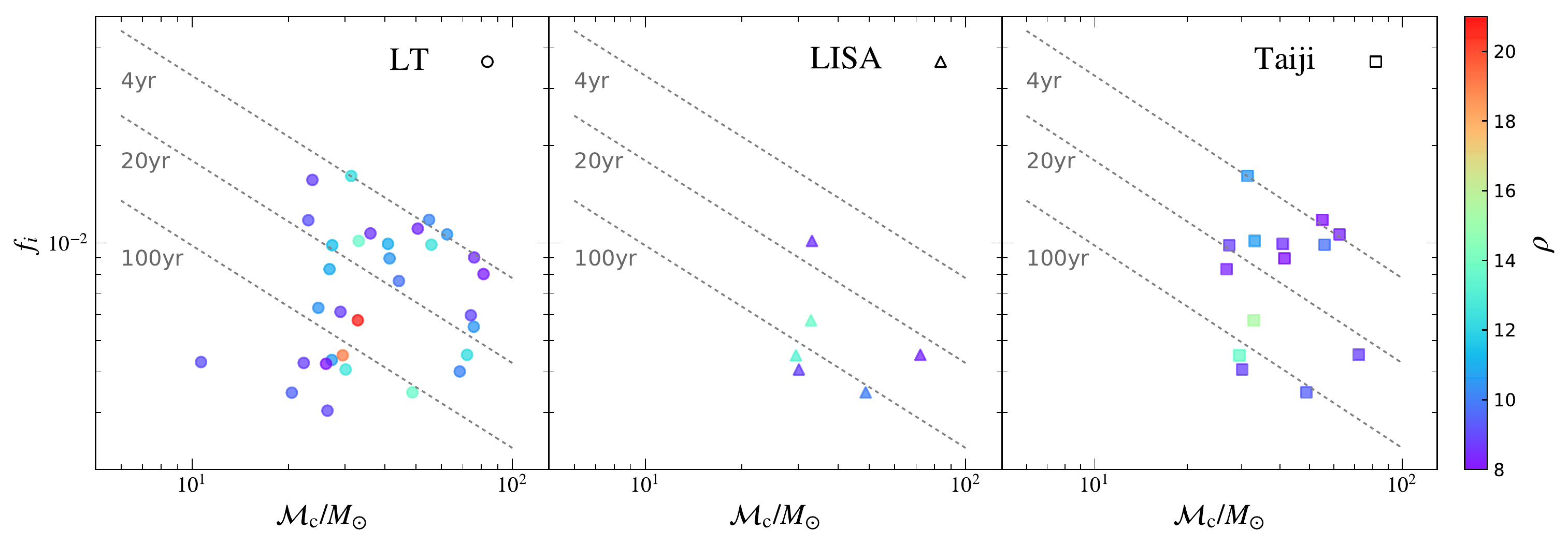}
\caption{
The initial observed frequencies and redshifted chirp masses of mock stellar BBHs resulting from a realization that can be detected by the LISA-Taiji (LT) network, LISA, and Taiji, respectively. Left, middle, and right panels show those mock stellar BBHs being detected by the LISA-Taiji network with SNR $\varrho_{\rm LT} \geq 8$, by LISA with $\varrho_{\rm L}\geq 8$, and by Taiji with $\varrho_{\rm T}\geq 8$, respectively, over an observation period of $4$ years. The color of each symbol marks the SNR of the source as indicated by the side colorbar. 
}
 \label{fig:frate4}
\end{figure}

Assuming a fixed observation period of $4$ years, we find that LISA, Taiji, or LISA-Taiji may detect $3$-$15$, $5$-$24$, or $15-58$ stellar BBHs by considering of the uncertainty in the local merger rate density constrained by LIGO/Virgo observations.  If adopting $\frac{d^2N}{dV dt}(z=0)  =19.1 \gpcyr$, then the expected number of ``detectable'' stellar BBHs is  $6$, $14$, and  $32$ for LISA, Taiji, and LISA-Taiji, respectively, according to the $50$th realization generated for this model. Fig.~\ref{fig:frate4} shows the  distributions of those ``detectable'' stellar BBHs in the observed initial frequency versus redshifted chirp mass plane. As seen from this figure, none of the $6$ stellar BBHs detected by LISA can merger within $4$ years and only $1$ can merger within $20$ years; $3$ of the $14$ BBHs detected by Taiji can evolve to high-frequency band and merger within $4$ years, and $8$ of them can merger within $20$ years; $3$/$17$  among the $32$ stellar BBHs detected by LISA-Taiji can merger within $4$/$10$ years; $9$ stellar BBHs have coalesecence time longer than $100$ years. The number of ``detectable'' stellar BBHs increases by a factor of $\sim 2-5$ comparing with those expected by Taiji or LISA individually.

\begin{figure}
\centering
\includegraphics[width=0.7\textwidth]{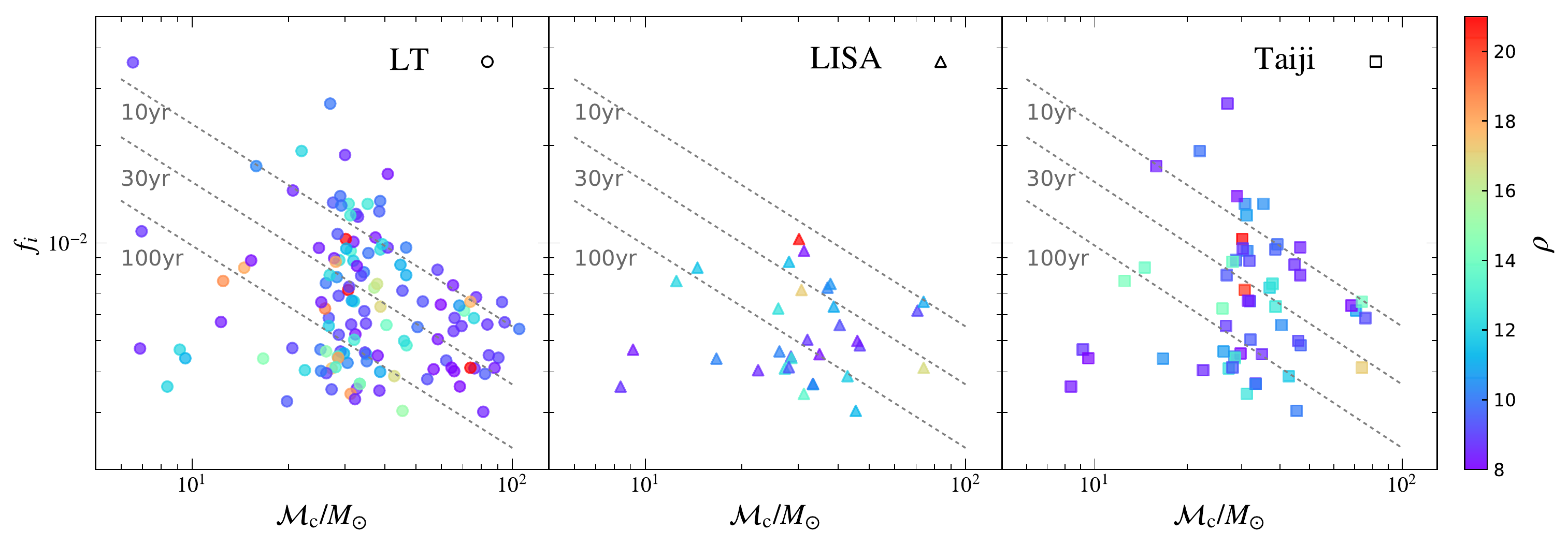}
\caption{Legend is the same as that for Fig.~\ref{fig:frate4}, except for an observation period of $10$ years.}
\label{fig:frate10}
\end{figure}

If extending the observation time to $10$ years, then  the expected number of ``detectable'' stellar BBHs increases to  $33$, $56$, or $132$ for LISA, Taiji, or LISA-Taiji according to $50$th realization from the model with $\frac{d^2N}{dV dt}(z=0)  =19.1 \gpcyr$. Fig.~\ref{fig:frate10} shows the ``detectable'' mock stellar BBHs for this case in the observed initial frequency versus redshifted chirp mass plane. As seen from this figure, LISA can detect many more stellar BBHs than the case shown in Fig.~\ref{fig:frate4} but still none of them can merge within $10$ years; Taiji can detect about $8$ stellar BBHs that can merge  within $10$ years; while LISA-Taiji can detect $24$ stellar BBHs that can merge within $10$ years. Apparently, the network by combining LISA and Taiji may improves subtantially the detection rate of stellar BBHs in the low-frequency band and enable the multi-band studies of a substantial number of stellar BBHs, which greatly enhances the potential of multi-band observations.

\begin{figure}
\centering
\includegraphics[width=0.7\textwidth]{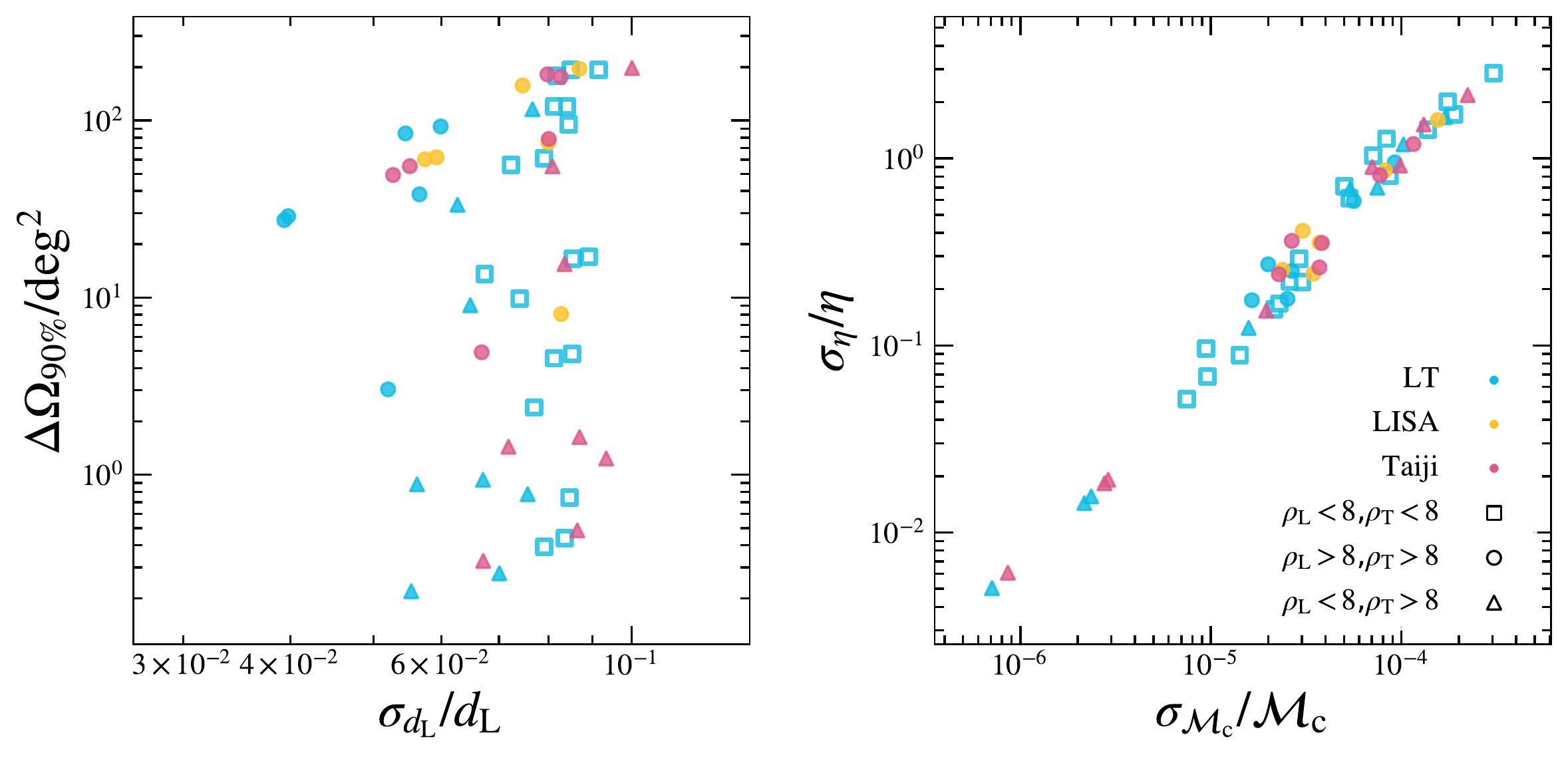}
\caption{
Estimates for the errors of parameter measurements for mock stellar BBHs resulting from a realization detected by LISA, Taiji, and the LISA-Taiji (LT) network, respectively (shown in Fig.~\ref{fig:frate4}). Left panel shows the sky localization errors ($90\%$ confidence level) and the relative errors of luminosity distance measurements for those BBHs. Right panel shows the estimated relative errors for the redshifted chirp masses and symmetric mass ratios. Yellow, red, and cyan symbols represent those sources detected by LISA, Taiji, and the LISA-Taiji network, respectively. Circles, triangles, and squares represent those sources detected by LISA only (totally six sources), by Taiji only but not LISA (totally eight sources) , and only by the LISA-Taiji network but not LISA or Taiji (totally eighteen sources), respectively.
}
\label{fig:fpara4}
\end{figure}

\begin{figure}
\centering
\includegraphics[width=0.7\textwidth]{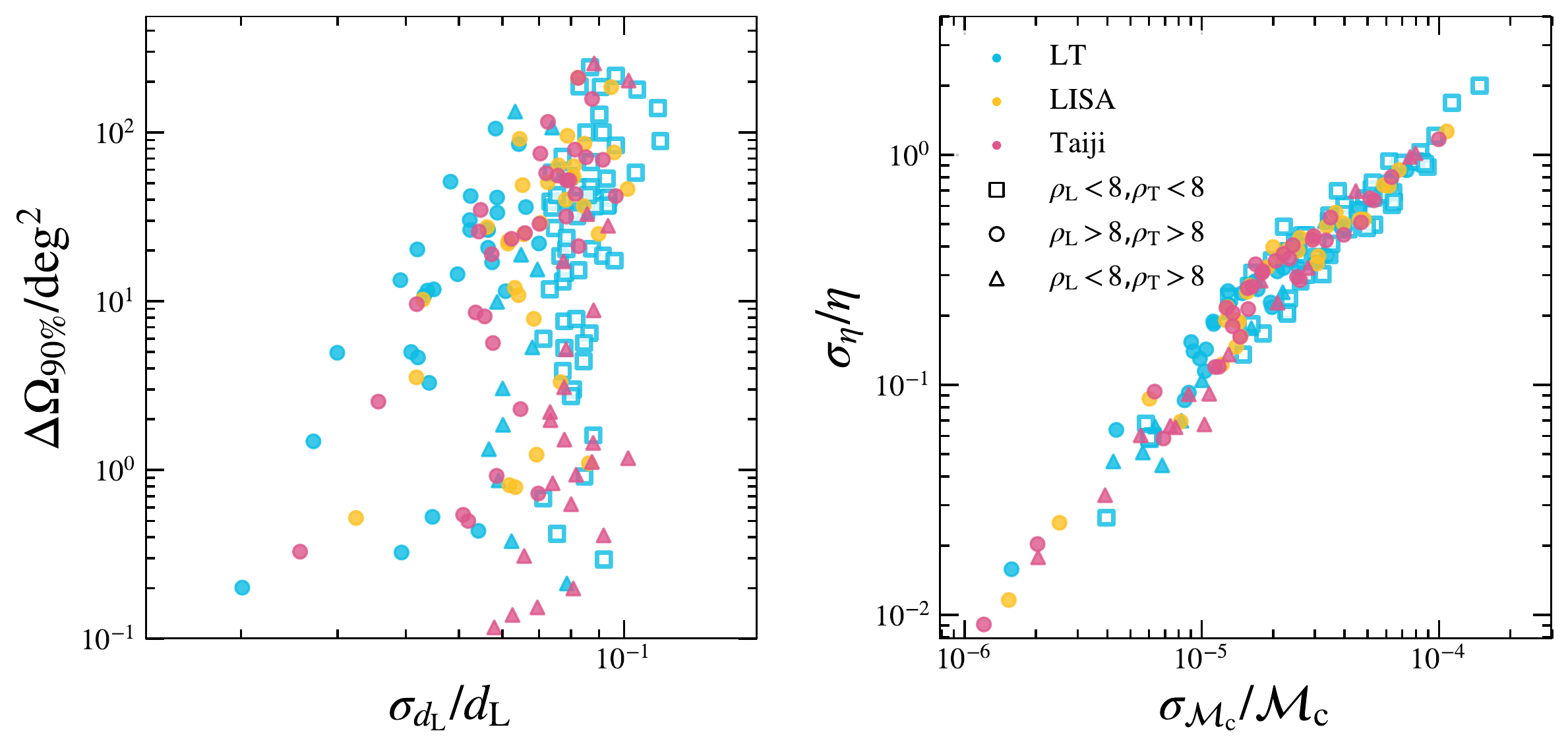}
\caption{Legend is the same as that for Fig.~\ref{fig:fpara4}, except for an observation period of $10$ years.}
\label{fig:fpara10}
\end{figure}

We also calculate the parameter estimation accuracies for those ``detectable'' stellar BBHs by LISA, Taiji, and LISA-Taiji, via the Fisher Information Matrix method. Figs.~\ref{fig:fpara4} and \ref{fig:fpara10} show the sky localization areas and the relative errors of luminosity distance measurements for BBHs detected by LISA, Taiji, and LISA-Taiji, via the observations with a period of $4$ and $10$ years, respectively. Assuming an observation period of $4$ years (see Fig.~\ref{fig:fpara4}), LISA or Taiji may localize the ``detectable'' BBHs in sky areas ranging from $\sim8.1$ to $190$\,deg$^2$ or from $\sim 0.33$ to $200$\,deg$^2$. LT may localize the ``detectable'' BBHs in sky areas in the range of $0.22$-$190$\,deg$^2$. For the common sources detected also by either LISA or Taiji,  the localizations by LISA-Taiji are a factor of $\sim 3$ or $\sim 2$ better than those obtained by using LISA or Taiji data only. The relative errors of luminosity distances detected by LISA or Taiji both ranges from $\sim 0.05$ to $0.1$. LISA-Taiji measure the luminosity distances with relative errors in the range of $0.04$-$0.09$. For those common sources detected also by either LISA or Taiji, LISA-Taiji gives the luminosity distance measurements slightly better than those obtained by using LISA or Taiji data only and the improvement factor is $\sim 1.3$. Similarly, LISA-Taiji could also help improving the measurement accuracy for $\mc$ and $\eta$ as well. LISA or Taiji could measure $\mc$ with relative errors in the range of $2.4\times10^{-5}$-$1.5\times10^{-4}$ or $8.6\times10^{-7}$-$2.2\times10^{-4}$ and $\eta$ with relative errors in the range of $0.24$-$1.6$ or $0.0061$-$2.2$.  LISA-Taiji could measure the redshifted chirp mass with relative errors in the range of $7.1\times 10^{-7}$-$3.0\times10^{-4}$ and symmetric mass ratio with relative errors in the range of $0.005$-$2.8$. If alternatively assuming an observation period of $10$ years (Fig.~\ref{fig:fpara10}), LISA, Taiji and LISA-Taiji may localize the detectable BBHs in sky areas in the range of $0.52$-$210$\,deg$^2$, $0.12$-$260$\,deg$^2$, and $0.07$-$240$\,deg$^2$,respectively; LISA, Taiji, and LISA-Taiji may measure the luminosity distances with relative errors in the range of $0.032$-$0.10$, $0.026$-$0.10$,  or $0.028$-$0.12$, measure the redshifted chirp masses with relative errors in the range of $1.5\times10^{-6}$-$1.1\times10^{-4}$, $3.7\times10^{-7}$-$1.0\times10^{-4}$, or $1.7\times10^{-7}$-$1.5\times10^{-4}$, and measure the symmetric mass ratios with relative errors in the range of $0.011$-$1.3$, $0.0027$-$1.2$, or $0.0016$-$2.0$. The reason for the slight improvements in localization and parameter estimations is mainly due to the enhancement of the SNRs for those sources.

In conclusion, the joint observations of LISA and Taiji may detect about several tens to a hundred stellar BBHs within an observation period of $4-10$ years, a factor of $2-3$ times of those obtained by using the observations by LISA or Taiji only in the same period.  Among these BBHs detected by LISA-Taiji, a few to twenty BBHs can merger within the observation period, which enables the multi-band GW observations and may realize the potential of multi-band GW observations in pre-warning of the GW and electromagnetic observations, understanding the formation mechanism of stellar BBHs, testing the gravity theory, etc. Combining the observations of LISA and Taiji may also improve the localization and measurement accuratenecss  of the system parameters, though the improvement is not as much as that for massive binary black holes.   


\section{Alternative networks \label{sec:an}}

Beyond the fiducial LISA-Taiji and LISA-TianQin networks discussed in the previous sections, alternative networks could also be formed by planned detectors, especially for the LISA-Taiji network.
The TianQin mission will deploy three spacecraft in a geocentric orbit with a semi-major axis of $\sim 10^5$ km, and the constellation will face to a binary white dwarf, J0806.3+1527, as a reference source \cite{luo2016tianqin}. Therefore, the fiducial case should be solo option for the LISA-TianQin or Taiji-TianQin network.
The LISA/Taiji missions will employ an Earth-like heliocentric orbit, and its constellation will be located at $\sim 20^\circ$ ($\sim 5.2 \times 10^7$ km) from the Earth \cite{LISA:2017pwj,Hu:2017mde}. This separation is a practical compromise between reducing Earth's gravitational perturbation and the launch vehicle, telemetry capabilities \cite{LISA2000}. The constellation plane is designed to be $60^\circ$ inclined with the ecliptic plane to achieve stable interferometer arms in the Clohessy-Wiltshire framework \cite{Dhurandhar:2004rv}.

\begin{figure}[htb]
\includegraphics[width=0.48\textwidth]{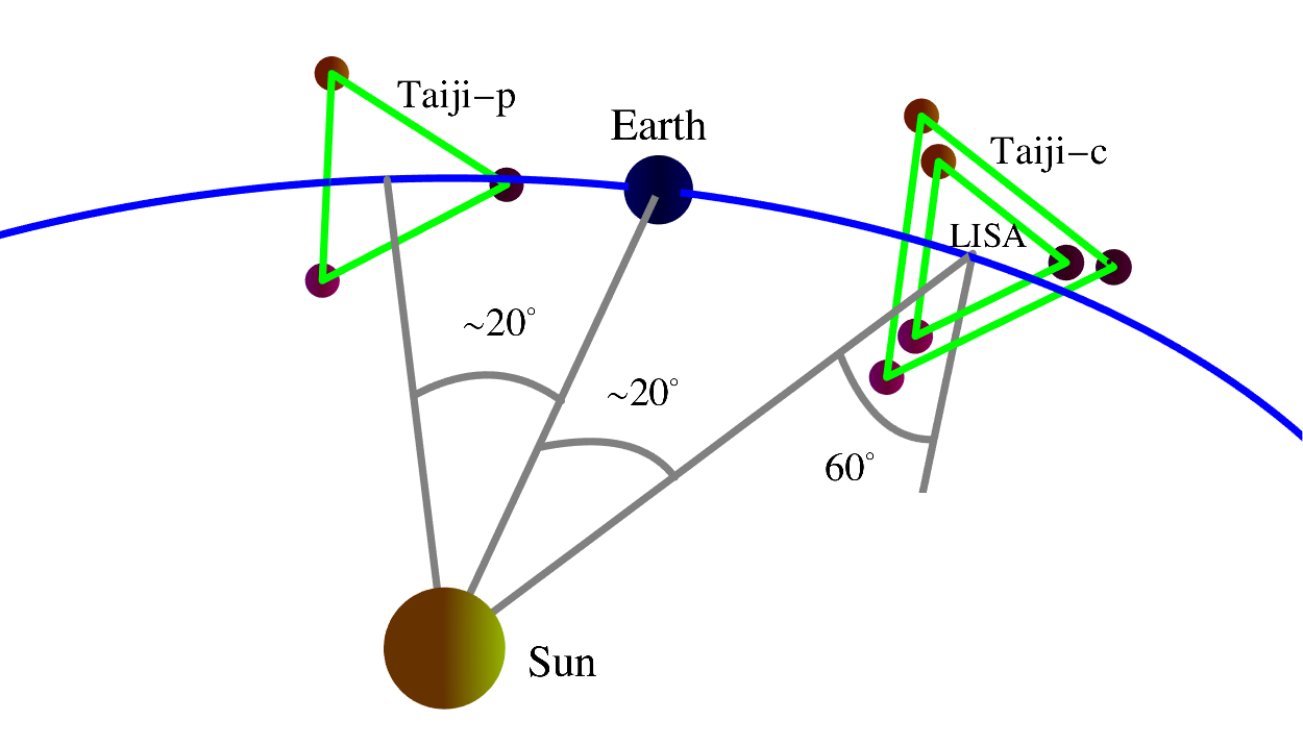}
\includegraphics[width=0.48\textwidth]{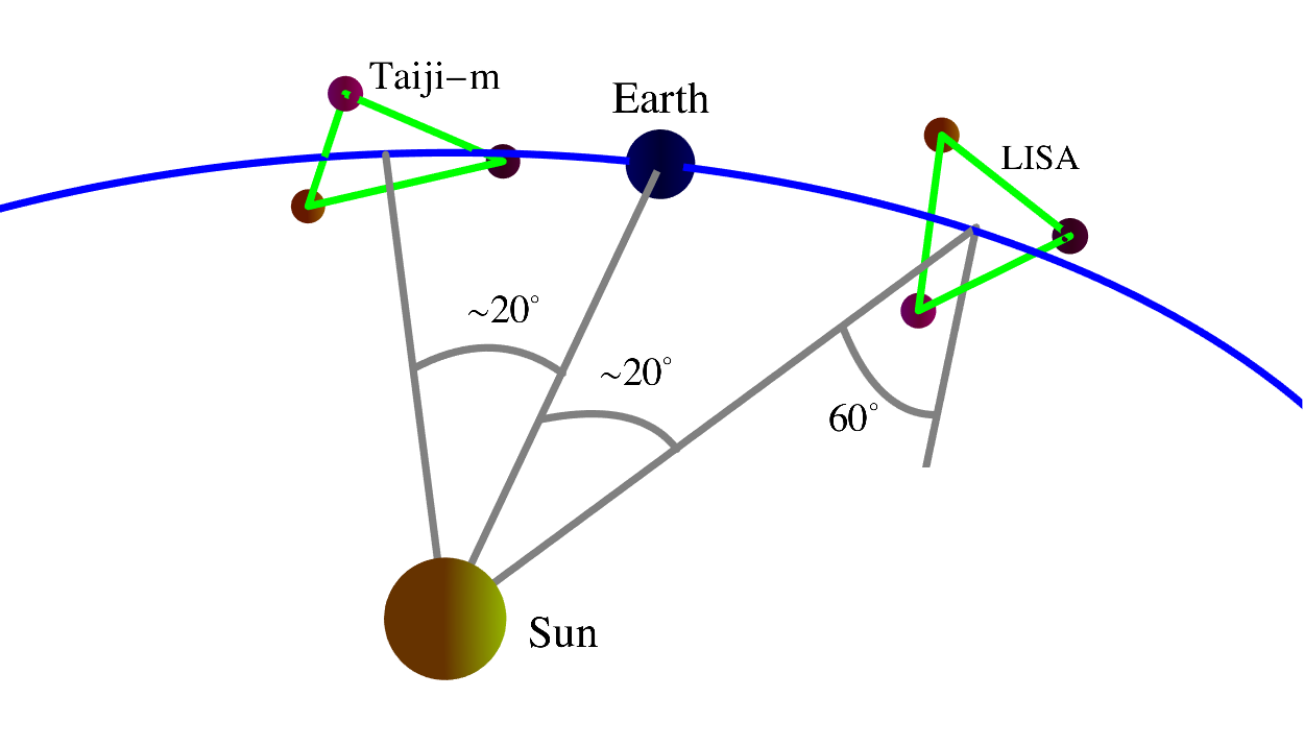}
\caption{\label{fig:alternative_LISA_TAIJI} Diagrams of the LISA and alternative Taiji mission orbital configurations. The left plot shows the LISA (trailing the Earth by $\sim$20$^\circ$ and $+60^\circ$ inclined with respect to the ecliptic plane), Taiji-p (leading the Earth by $\sim$20$^\circ$ and $+60^\circ$ inclined), and Taiji-c (colocated and coplanar with LISA). The right panel shows the LISA and Taiji-m (leading the Earth by $\sim$20$^\circ$ and $-60^\circ$ inclined). The angle between the LISA and Taiji-m constellation planes is $\sim71^\circ$, and the angle between LISA and Taiji-p formation planes is $\sim34.5^\circ$ (recreated from \cite{Wang:alternative1}).
}
\end{figure}

The LISA will trail the Earth by $20^\circ$, and the formation plane is $+60^\circ$ inclined. Since the Taiji orbit configuration is still not fully determined, its location and orientation could be adjusted to form different networks with LISA. Three alternative LISA-Taiji networks are introduced by altering the location and orientation of the Taiji in \cite{Wang:alternative1,Wang:alternative2}, 
\begin{itemize}
    \item[(a)] Taiji-p (fiducial case), which leads the Earth by $\sim 20^\circ$ and the formation plane is $+60^\circ$ inclined as the LISA.
    \item[(b)] Taiji-c, which is co-located and coplanar with the LISA.
    \item[(c)] Taiji-m, which leads the Earth by $\sim 20^\circ$ and the formation plane is $-60^\circ$ inclined.
\end{itemize}
And the three Taiji orbital configurations are shown in Fig.~\ref{fig:alternative_LISA_TAIJI}. There also could be another Taiji orbit which trails the Earth by $\sim 20^\circ$ as LISA but with $-60^\circ$ orientation. And this case is ignored since it loses both the long baseline pros for the compact binary observation and full correlation with LISA for the SGWB observation.
The launch budget for these three Taiji deployments should be identical/comparable, and the observation achievement should not be (statistically) different from each other. Without considering the restriction of $20^\circ$ separation from the Earth, \cite{shuman2022massive} examined more exhaustive Taiji orbit deployments to form the LISA-Taiji networks.

\begin{figure}[htb]
\includegraphics[width=0.48\textwidth]{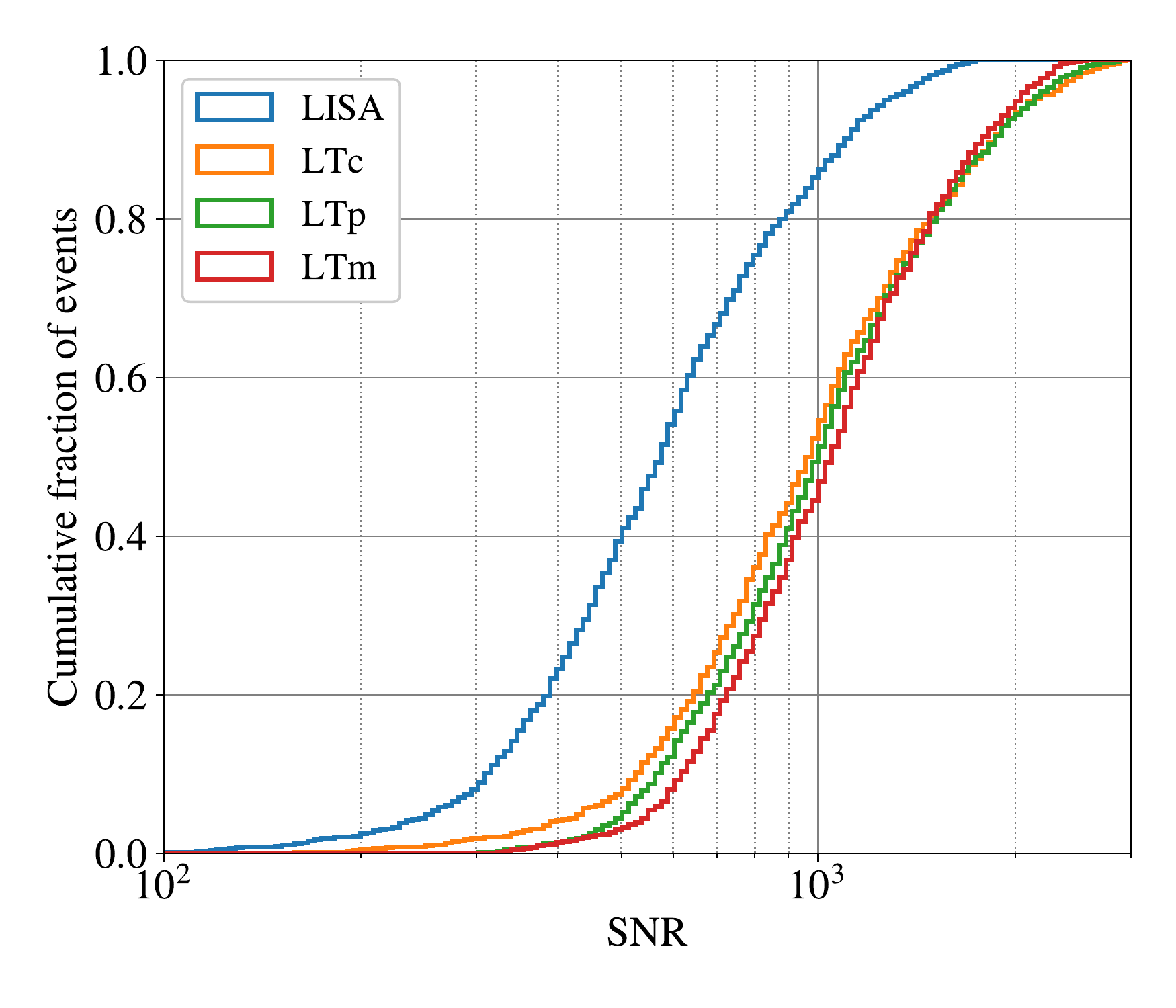}
\includegraphics[width=0.48\textwidth]{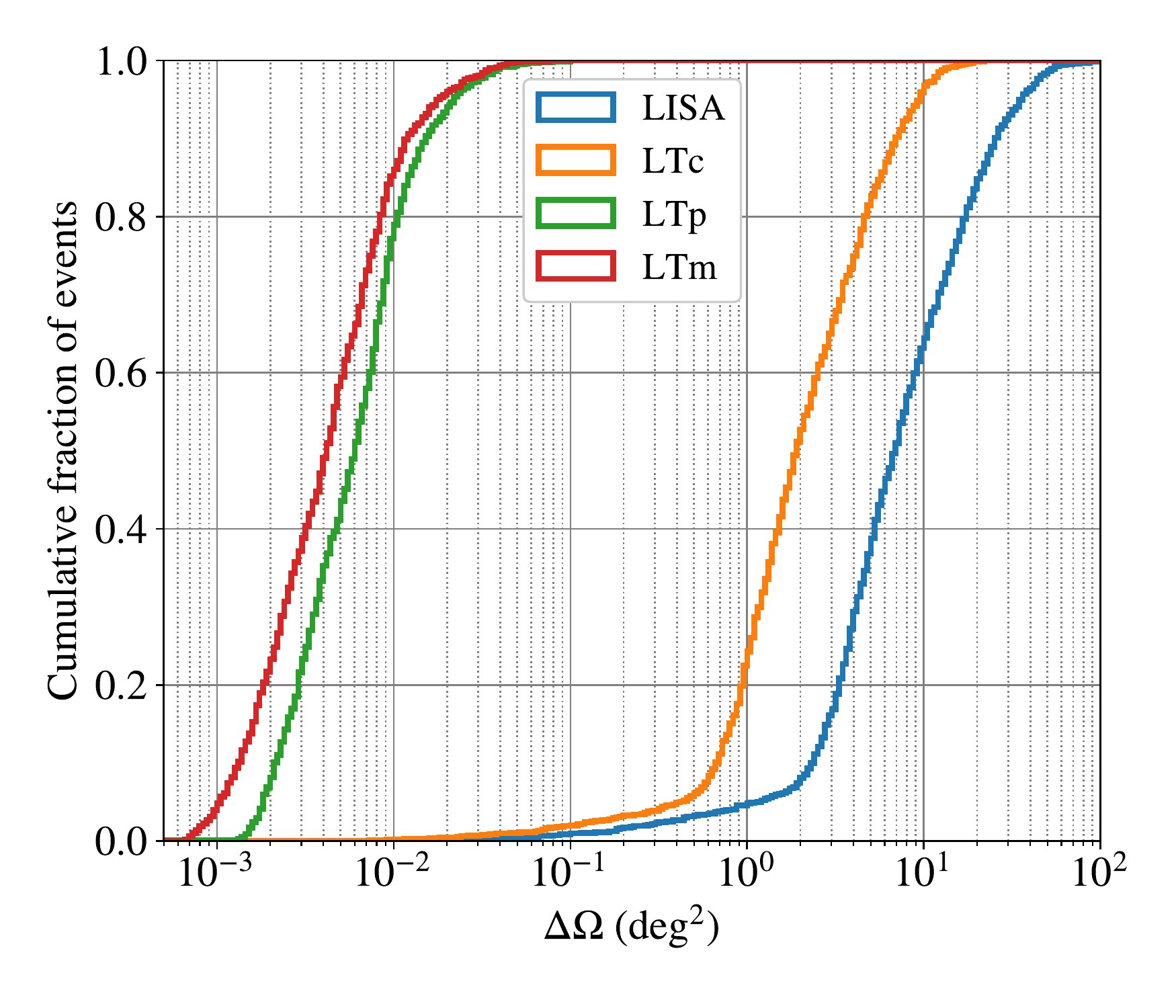}
\caption{\label{fig:SNR_SkyLoc_LISA_TAIJI} The cumulative histograms of SNR (left panel) and angular resolution of the sky localization (right panel) from LISA and three LISA-Taiji networks (LTc: LISA with Taiji-c, LTp: LISA with Taiji-p, LTm: LISA with Taiji-m) for the massive BBH mergers (which masses $m_1 = 10^5 M_\odot, m_2= m_1/3$, at redshift $z=2$, and their sky directions are random) (figures from \cite{Wang:alternative1}).
}
\end{figure}

Compared to a single detector, the joint observation will promote the SNR of GW detections by a factor of $\sim \sqrt{N_\mathrm{det}}$, where $N_\mathrm{det}$ is the number of detectors (with comparable sensitivities), and the joint SNR will be $\rho^2_\mathrm{joint} = \sum^{N_\mathrm{det}}_{i=1} \rho^2_i$. 
Ref.~\cite{Wang:alternative1} evaluated the SNRs and angular resolutions of sky localization from three LISA-Taiji networks by simulating massive BBH population with fixed masses ($m_1 = 10^5 M_\odot, m_2= m_1/3$) at redshift $z=2$ and randomly and uniformly distributed sky directions, and the results are shown in Fig.~\ref{fig:SNR_SkyLoc_LISA_TAIJI}. Considering the Taiji is more sensitive than the LISA, the SNRs (in the last month before coalescences) obtained by the networks are more than $\sim\sqrt{2}$ times higher than the single LISA as shown in the left panel. The distribution of SNRs from the joint LISA and Taiji-m observations are more concentrated compared to the other two networks, and the reason is that the Taiji-m could compensate for the LISA's insensitive direction, and their joint response function would be more isotropic to all sky directions. 

On the other hand, although the SNRs from the three networks are comparable, their capabilities of sky localization are rather different \cite{Wang:alternative1,shuman2022massive}. The first reason is that the larger separation between LISA and Taiji-p/-m could improve the resolution of sky localization, and the second reason is that the different orientations will have different antenna patterns which yield different impacts on the direction determinations. Benefiting from these two factors, as shown in the right panel of Fig.~\ref{fig:SNR_SkyLoc_LISA_TAIJI}, the joint LISA and Taiji-m observations have the best performance on the sky localization for the sources and could improve the accuracy by a factor of $\sim$3 orders compared to the LISA. The capability of LISA-Taiji-p is slightly worse than LISA-Taiji-m due to the suboptimal orientations, and the LISA-Taiji-c only improves the sky localization by a factor of $\sim$2 attributed to the increase of detectors.

By employing the LISA-TianQin and alternative LISA-Taiji networks, \cite{shuman2022massive} investigated the joint observations of the massive BBH from light seed (popIII) and heavy seed (Q3\_nodelay) which simulated in \cite{Klein:2015hvg}. The capability of the LISA-Taiji-p/LISA-Taiji-m would be comparable to the LISA-TianQin to localize the BBH for the light seed scenario, and its performance become better than the LISA-TianQin for the heavy seed case because  Taiji has a better sensitivity than the TianQin in the lower frequencies. 

Beyond tensor polarization observations, the polarization of scalar and vector modes could also be better observed/constrained by the LISA-Taiji network. By employing the parametrized post-Einsteinian (ppE) formalism developed in \cite{ppe2009PhRvD..80l2003Y,yunes2012PhRvD..86b2004C}, the determinations of the ppE parameters are examined in \cite{Wang:2021mou,Wang:alternative1}. And the results show that LISA-Taiji-m would have the best performance in three alternative networks, and it could refine the amplitudes of scalar and vector polarizations by $\sim$10 times compared to the single LISA.

\begin{figure}[htb]
\includegraphics[width=0.48\textwidth]{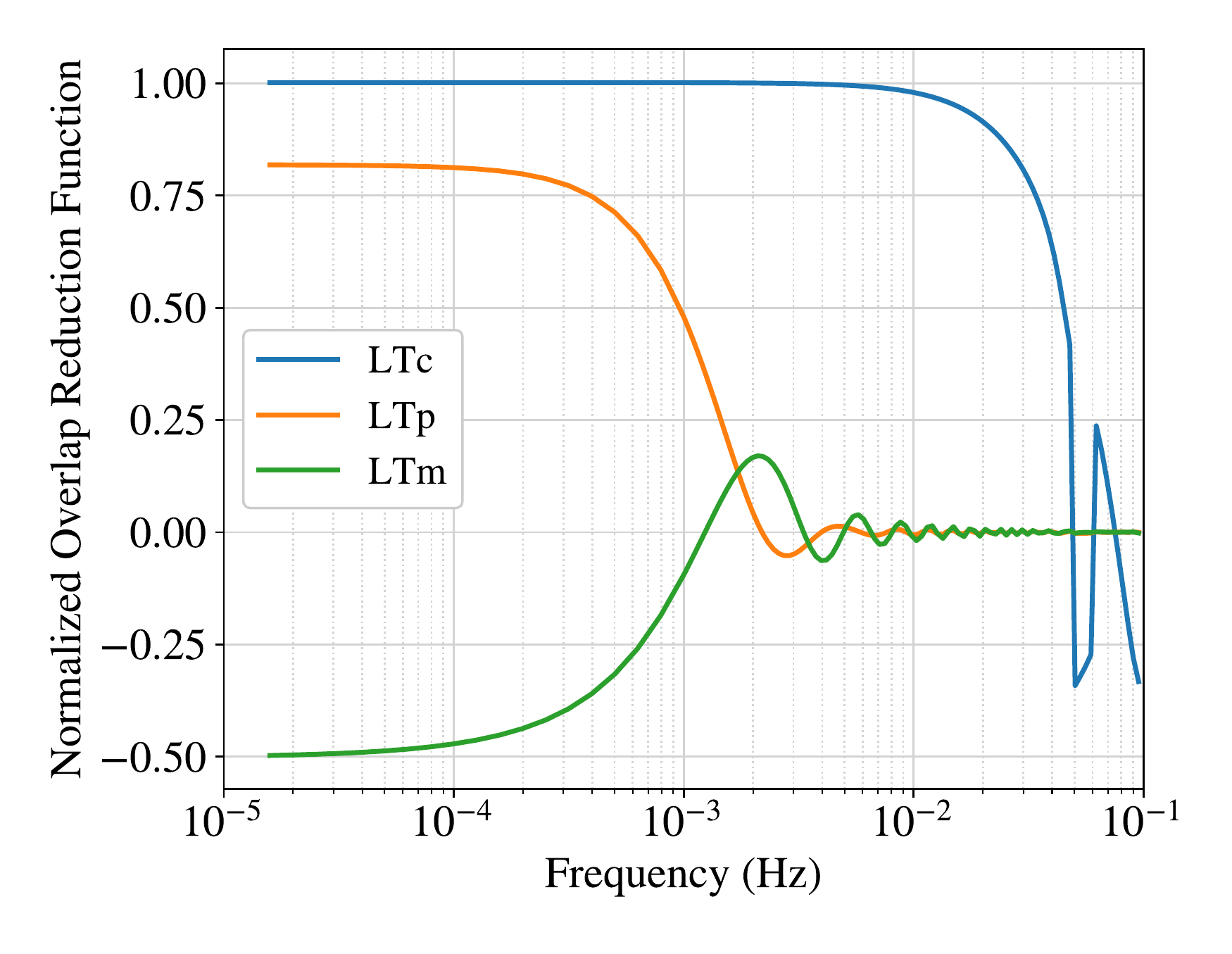}
\includegraphics[width=0.46\textwidth]{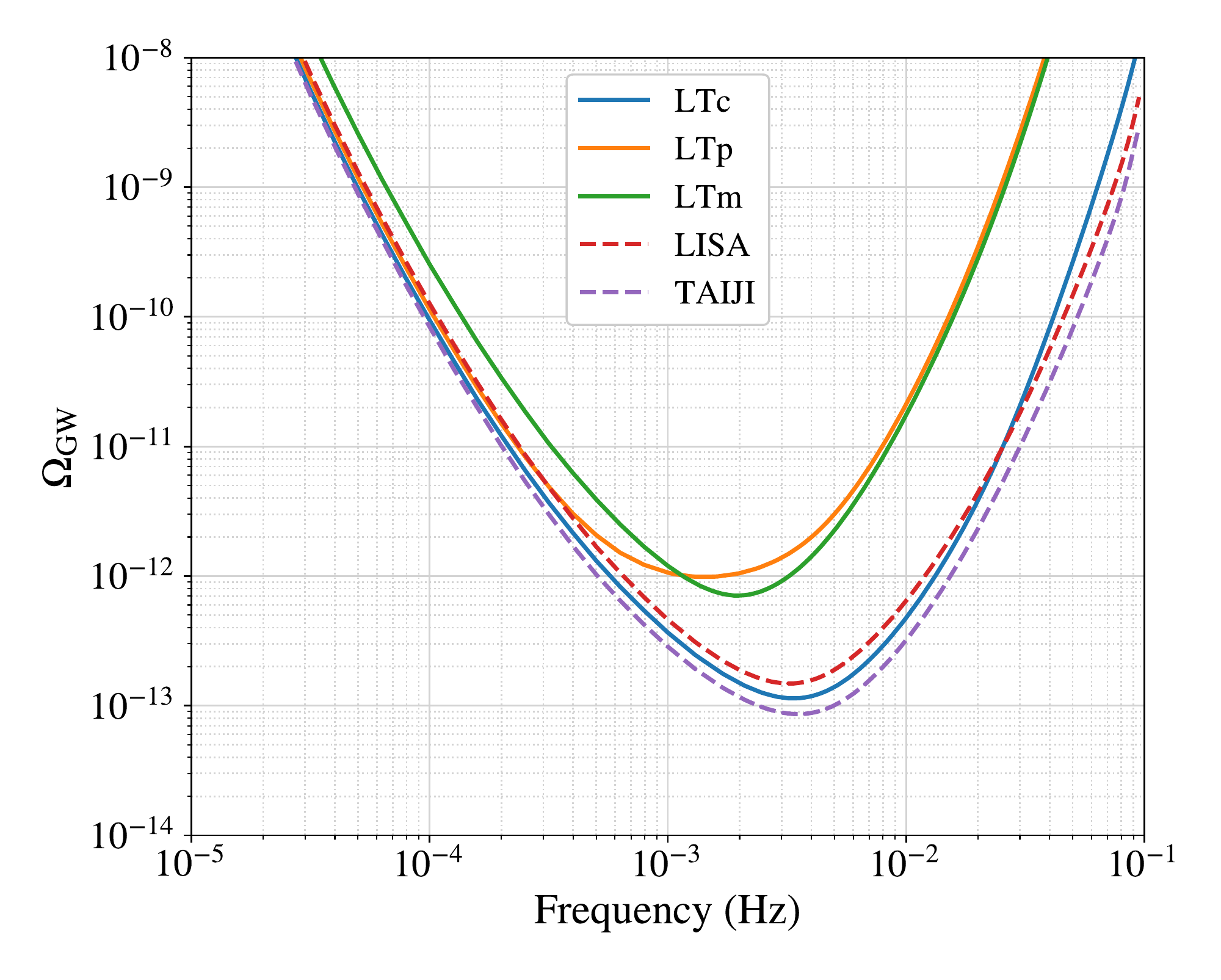}
\caption{\label{fig:ORF_LISA_TAIJI} The overlap reduction functions for tensor polarization (left) and power-law integrated sensitivities (right) of three LISA-Taiji networks (LTc: LISA with Taiji-c, LTp: LISA with Taiji-p, LTm: LISA with Taiji-m). The overlap reduction function is calculated for the tensor $+$ and $\times$ polarizations, and the sensitivities in the right plot is obtained by assuming the SNR threshold $\rho_{th}=10$ in three-year observation (figures from \cite{Wang:alternative1,Wang:alternative2}).
}
\end{figure}

The SGWB is another targeting GW source for space missions, it may be identified from the cross-correlation between two detectors by assuming the noises are not correlated between the two detectors. The correlation between detector $a$ and $b$ could be quantified by overlap reduction function (ORF) as discussed in Sec.~\ref{sec:test_pv}.
The normalized ORFs of three LISA-Taiji networks are shown in the left panel of Fig.~\ref{fig:ORF_LISA_TAIJI}. As we can read from the plot, the ORF of LISA-Taiji-c is unity for frequencies lower than $\sim$10 mHz since two detectors are coplanar at the same location. It means that SGWB observed by LISA and Taiji-c will be fully correlated, and LISA-Taiji-c will be an optimal network for the SGWB identification. Due to the separation, $D=1 \times 10^8$ km, between LISA and Taiji-p/-m, their ORFs quickly approach zero around a critical frequency $f_c = c/(2D) \simeq 1.5$ mHz and oscillate and decay with the increase of frequency \cite{Romano:2016dpx}. The amplitude of ORF from LISA-Taiji-m is lower than LISA-Taiji-p in frequencies lower than the critical frequency, and it indicates the former would have the worse capability to observe the SGWB than the latter in this band.

The power-law integrated sensitivity could be employed to illustrate the detectability of the power-law SGWB signal \cite{Thrane:2013oya}, and the sensitivities of three network configurations are shown in the right panel of Fig.~\ref{fig:ORF_LISA_TAIJI} \cite{Wang:alternative2}. As we can expect, the LISA-Taiji-c is optimal for the SGWB observation. For the other two networks, the LISA-Taiji-p is more sensitive than LISA-Taiji-m for frequencies lower than $\sim$1 mHz, and the LISA-Taiji-m becomes more sensitive in the frequency band of $\sim$[1, 8] mHz which is also the most sensitive band of the detectors. Their sensitivities converge in the higher frequency range. Based on their pros and cons at different frequencies, the performance of LISA-Taiji-p and LISA-Taiji-m could be moderately/slightly different for a specific sensible SGWB as investigated in \cite{Wang:alternative2}.

The BBO and DECIGO are planned to place three constellations on the Earth-like heliocentric orbit and observe the GW in the deci-Hz frequency band \cite{BBO:2005,DECIGO:2006}. The separation between two constellations will be $\sqrt{3}$ AU, and the formation plane will be $60^\circ$ inclined with respect to the ecliptic plane. Benefiting from the long baselines and synergetic antenna pattern, the parameters of the compact binaries could be well resolved \cite{BBO:2005}.

\section{Conclusions and prospects \label{sec:cp}}
The era of mHz GW detections are expected to begin in the 2030s, when multiple space-based detectors would observe various sources in the mHz frequency band simultaneously. In this work, we discuss the benefits of joint observations from space detector networks. These will offer a number of advantages, including 
\begin{itemize}
    \item More precise localization of GW sources, which will be essential for understanding their astrophysical origins.
    \item Improved tests of relativistic gravity, including measurements of GW polarization and the possible rotation of the cosmic polarization vector.
    \item Better determination of the galactic background of GWs, which will provide insights into the formation and evolution of our galaxy.
    \item An increase in the event rate of GW detections, including both stellar-mass black hole mergers and massive black hole mergers.
\end{itemize}
The alternative LISA-Taiji networks are examined to determine the best configuration for maximizing the scientific benefits. The Taiji-m configuration is particalarly promising, and it could be a competitive configuraiton as the fiducial Taiji-p case. And more specific investigations should be performed before the Taiji team determines the final orbital configuration.

The joint observation of space-based GW network would significantly improve the determinations of parameters from massive BBHs. Benefiting from their large separation and complementary orientation, the parameters of sources would be measured with significant improvement compared to a single detector, especially for the sky localization. The individual detector orbits around the Sun/Earth which is equivalent to a multitude of detectors at different positions and times. This would allow for the identifications of (quasi-)periodical sources' directions. The early and quick localization from the network will be crucial for multi-messenger observation, such as those performed by the ground-based advanced LIGO and advanced Virgo observations.

The joint network would also be decisive in distinguishing SGWB from the detection noises. The SGWB encodes important information about the early Universe and binary population. It would be difficult for a single detector to separate the stochastic signals from the stochastic noise, especially for an unforeseen spectrum. With joint observations of independent detectors, similar to the SGWB search employed by the ground-based interferometers, the common signals could be extracted from the correlation between two detector's data. And the physical properties, for instance, the parity violation, could be tested with the two detectors.

A network of GW detectors operating in the same frequency band can increase the efficiency of detections. Moreover, multi-band observation, either at the same time or at different times, can also increase the number of detectable events. If the phenomenological model or theoretical template is robust, observations crossing different frequency bands would tighten the parameter estimation even more effectively. For example, with a connected binary inspiral template through mHz, deci-Hz and deca-Hz frequencies, a network of space GW detectors of LISA, Taiji, AMIGO, ET and CE will be able to enhance the parameter estimation by two orders of magnitude \cite{Zhao:2023}. These enhancements will strengthen the distinguishability of various GW source models, the precision of determination of cosmological models and Hubble constant, and the co-evolution of star formation, black holes and galaxies.

In addition to the mHz missions, AU-sized arm length interferometers have been proposed to observe the GW in the sub-mHz to $\mu$Hz band which includes the ASTROD-GW \cite[and references therein]{Ni:2012eh,Ni:2016wcv} , Folkner’s mission \cite{Baker:2019pnp} $\mu$-Aries \cite{Sesana:2019vho}, and LISAmax \cite{Martens:2023mgm}. The technologies for these missions are considered to be as ready as that for the LISA. However, their sensitivity will be limited not only by instrumental noises but also confusion foreground from binaries in our galaxy \cite{Wang:2023jct}. Even so, the AU-size interferometers could still be more sensitive than the Gm-sized detectors by a ratio of their arm lengths. With the low-frequency GW detector(s), the massive BBHs could be routinely observed, and their distances and redshifts would be unambiguously determined with multi-messenger observations. The cosmic evolution achieved from the observation in these frequencies will give input to the issues of dark energy and dark matter, as well as their evolution.

\paragraph{Acknowledgments.} RGC is supported in part by the National Natural Science Foundation of China Grant No. 11821505. ZKG is supported in part by the National Key Research and Development Program of China Grant No.2020YFC2201501, in part by the National Natural Science Foundation of China Grants No. 12075297 and No. 12235019.
BH is supported in part by the National Key R\&D Program of China No. 2021YFC2203001.
CL is supported by the National Natural Science Foundation of China Grant No. 12147132.
YJL is partly supported by the National Key Program for Science and Technology Research and Development (grant nos. 2020YFC2201400, 2022YFC2205201). 
WTN is supported in part by the National Key R\&D Program of China under Grant No. 2021YFC2201901.
NS is supported in part by JSPS Kakenhi Grant-in-Aid for  Scientific Research Numbers 17H06358, 19K03870 and 23K03385.
GW is supported by the National Key R\&D Program of China under Grant No. 2021YFC2201903, and NSFC No. 12003059.

\newpage
\bibliographystyle{unsrt}
\bibliography{network}

\begin{thebibliography}{100}

\bibitem{Schutz:1986gp}
Bernard~F. Schutz.
\newblock {Determining the Hubble Constant from Gravitational Wave
  Observations}.
\newblock {\em Nature}, 323:310--311, 1986.

\bibitem{LIGOScientific:2016aoc}
B.~P. Abbott et~al.
\newblock {Observation of Gravitational Waves from a Binary Black Hole Merger}.
\newblock {\em Phys. Rev. Lett.}, 116(6):061102, 2016.

\bibitem{LIGOScientific:2017ycc}
B.~P. Abbott et~al.
\newblock {GW170814: A Three-Detector Observation of Gravitational Waves from a
  Binary Black Hole Coalescence}.
\newblock {\em Phys. Rev. Lett.}, 119(14):141101, 2017.

\bibitem{LIGOScientific:2017vwq}
B.~P. Abbott et~al.
\newblock {GW170817: Observation of Gravitational Waves from a Binary Neutron
  Star Inspiral}.
\newblock {\em Phys. Rev. Lett.}, 119(16):161101, 2017.

\bibitem{Cutler:1997ta}
Curt Cutler.
\newblock {Angular resolution of the LISA gravitational wave detector}.
\newblock {\em Phys. Rev. D}, 57:7089--7102, 1998.

\bibitem{LISA:2017pwj}
Pau Amaro-Seoane, Heather Audley, Stanislav Babak, John Baker, Enrico Barausse,
  Peter Bender, Emanuele Berti, Pierre Binetruy, Michael Born, Daniele
  Bortoluzzi, et~al.
\newblock Laser interferometer space antenna.
\newblock {\em arXiv preprint arXiv:1702.00786}, 2017.

\bibitem{Ruan:2020smc}
Wen-Hong Ruan, Chang Liu, Zong-Kuan Guo, Yue-Liang Wu, and Rong-Gen Cai.
\newblock {The LISA-Taiji network}.
\newblock {\em Nature Astron.}, 4:108--109, 2020.

\bibitem{Ruan:2019tje}
Wen-Hong Ruan, Chang Liu, Zong-Kuan Guo, Yue-Liang Wu, and Rong-Gen Cai.
\newblock {The LISA-Taiji Network: Precision Localization of Coalescing Massive
  Black Hole Binaries}.
\newblock {\em Research}, 2021:6014164, 2021.

\bibitem{Schutz:2011tw}
Bernard~F. Schutz.
\newblock {Networks of gravitational wave detectors and three figures of
  merit}.
\newblock {\em Class. Quant. Grav.}, 28:125023, 2011.

\bibitem{Seto:2007tn}
Naoki Seto and Atsushi Taruya.
\newblock {Measuring a Parity Violation Signature in the Early Universe via
  Ground-based Laser Interferometers}.
\newblock {\em Phys. Rev. Lett.}, 99:121101, 2007.

\bibitem{Hu:2017mde}
Wen-Rui Hu and Yue-Liang Wu.
\newblock {The Taiji Program in Space for gravitational wave physics and the
  nature of gravity}.
\newblock {\em Natl. Sci. Rev.}, 4(5):685--686, 2017.

\bibitem{ruan2020taiji}
Wen-Hong Ruan, Zong-Kuan Guo, Rong-Gen Cai, and Yuan-Zhong Zhang.
\newblock Taiji program: Gravitational-wave sources.
\newblock {\em International Journal of Modern Physics A}, 35(17):2050075,
  2020.

\bibitem{luo2016tianqin}
Jun Luo, Li-Sheng Chen, Hui-Zong Duan, Yun-Gui Gong, Shoucun Hu, Jianghui Ji,
  Qi~Liu, Jianwei Mei, Vadim Milyukov, Mikhail Sazhin, et~al.
\newblock Tianqin: a space-borne gravitational wave detector.
\newblock {\em Classical and Quantum Gravity}, 33(3):035010, 2016.

\bibitem{hu2017science}
Yi-Ming Hu, Jianwei Mei, and Jun Luo.
\newblock Science prospects for space-borne gravitational-wave missions.
\newblock {\em National Science Review}, 4(5):683--684, 2017.

\bibitem{bogdanovic2022electromagnetic}
Tamara Bogdanovi{\'c}, M~Coleman Miller, and Laura Blecha.
\newblock Electromagnetic counterparts to massive black-hole mergers.
\newblock {\em Living Reviews in Relativity}, 25(1):3, 2022.

\bibitem{cutler1998angular}
Curt Cutler.
\newblock Angular resolution of the lisa gravitational wave detector.
\newblock {\em Physical Review D}, 57(12):7089, 1998.

\bibitem{moore2002angular}
Thomas~A Moore and Ronald~W Hellings.
\newblock Angular resolution of space-based gravitational wave detectors.
\newblock {\em Physical Review D}, 65(6):062001, 2002.

\bibitem{seto2004strong}
Naoki Seto.
\newblock Strong gravitational lensing and localization of merging massive
  black hole binaries with lisa.
\newblock {\em Physical Review D}, 69(2):022002, 2004.

\bibitem{lang2006measuring}
Ryan~N Lang and Scott~A Hughes.
\newblock Measuring coalescing massive binary black holes with gravitational
  waves: The impact of spin-induced precession.
\newblock {\em Physical Review D}, 74(12):122001, 2006.

\bibitem{lang2008localizing}
Ryan~N Lang and Scott~A Hughes.
\newblock Localizing coalescing massive black hole binaries with gravitational
  waves.
\newblock {\em The Astrophysical Journal}, 677(2):1184, 2008.

\bibitem{trias2008lisa}
Miquel Trias and Alicia~M Sintes.
\newblock Lisa observations of supermassive black holes: Parameter estimation
  using full post-newtonian inspiral waveforms.
\newblock {\em Physical Review D}, 77(2):024030, 2008.

\bibitem{thorpe2009lisa}
JI~Thorpe, ST~McWilliams, BJ~Kelly, RP~Fahey, K~Arnaud, and JG~Baker.
\newblock Lisa parameter estimation using numerical merger waveforms.
\newblock {\em Classical and Quantum Gravity}, 26(9):094026, 2009.

\bibitem{wang2020numerical}
Gang Wang, Wei-Tou Ni, Wen-Biao Han, Shu-Cheng Yang, and Xing-Yu Zhong.
\newblock Numerical simulation of sky localization for lisa-taiji joint
  observation.
\newblock {\em Physical Review D}, 102(2):024089, 2020.

\bibitem{zhang2021parameter}
Chunyu Zhang, Yungui Gong, Chao Zhang, et~al.
\newblock Parameter estimation for space-based gravitational wave detectors
  with ringdown signals.
\newblock {\em Physical Review D}, 104(8):083038, 2021.

\bibitem{shuman2022massive}
Kevin~J. Shuman and Neil~J. Cornish.
\newblock {Massive black hole binaries and where to find them with dual
  detector networks}.
\newblock {\em Phys. Rev. D}, 105(6):064055, 2022.

\bibitem{Wang:alternative1}
Gang Wang, Wei-Tou Ni, Wen-Biao Han, Peng Xu, and Ziren Luo.
\newblock {Alternative LISA-TAIJI networks}.
\newblock {\em Phys. Rev. D}, 104(2):024012, 2021.

\bibitem{Zhang:2021wwd}
Chunyu Zhang, Yungui Gong, and Chao Zhang.
\newblock {Source localizations with the network of space-based gravitational
  wave detectors}.
\newblock {\em Phys. Rev. D}, 106(2):024004, 2022.

\bibitem{korol2017prospects}
Valeriya Korol, Elena~M Rossi, Paul~J Groot, Gijs Nelemans, Silvia Toonen, and
  Anthony~GA Brown.
\newblock Prospects for detection of detached double white dwarf binaries with
  gaia, lsst and lisa.
\newblock {\em Monthly Notices of the Royal Astronomical Society},
  470(2):1894--1910, 2017.

\bibitem{lamberts2019predicting}
Astrid Lamberts, Sarah Blunt, Tyson~B. Littenberg, Shea Garrison-Kimmel, Thomas
  Kupfer, and Robyn~E. Sanderson.
\newblock {Predicting the LISA white dwarf binary population in the Milky Way
  with cosmological simulations}.
\newblock {\em Mon. Not. Roy. Astron. Soc.}, 490(4):5888--5903, 2019.

\bibitem{liu2023confusion}
Chang Liu, Wen-Hong Ruan, and Zong-Kuan Guo.
\newblock Confusion noise from galactic binaries for taiji.
\newblock {\em arXiv preprint arXiv:2301.02821}, 2023.

\bibitem{kupfer2018lisa}
T~Kupfer, V~Korol, S~Shah, G~Nelemans, TR~Marsh, G~Ramsay, PJ~Groot, DTH
  Steeghs, and EM~Rossi.
\newblock Lisa verification binaries with updated distances from gaia data
  release 2.
\newblock {\em Monthly Notices of the Royal Astronomical Society},
  480(1):302--309, 2018.

\bibitem{jiang2023sky}
Tong Jiang, Yungui Gong, and Xuchen Lu.
\newblock Sky localization of space-based detectors with time-delay
  interferometry.
\newblock {\em arXiv preprint arXiv:2301.05923}, 2023.

\bibitem{Zhang:2020hyx}
Chao Zhang, Yungui Gong, Hang Liu, Bin Wang, and Chunyu Zhang.
\newblock {Sky localization of space-based gravitational wave detectors}.
\newblock {\em Phys. Rev. D}, 103(10):103013, 2021.

\bibitem{Wang:2020dkc}
Renjie Wang, Wen-Hong Ruan, Qing Yang, Zong-Kuan Guo, Rong-Gen Cai, and Bin Hu.
\newblock {Hubble parameter estimation via dark sirens with the LISA-Taiji
  network}.
\newblock {\em Natl. Sci. Rev.}, 9(2):nwab054, 2022.

\bibitem{Wang:2021srv}
Ling-Feng Wang, Shang-Jie Jin, Jing-Fei Zhang, and Xin Zhang.
\newblock {Forecast for cosmological parameter estimation with
  gravitational-wave standard sirens from the LISA-Taiji network}.
\newblock {\em Sci. China Phys. Mech. Astron.}, 65(1):210411, 2022.

\bibitem{Yang:2021qge}
Tao Yang.
\newblock {Gravitational-Wave Detector Networks: Standard Sirens on Cosmology
  and Modified Gravity Theory}.
\newblock {\em JCAP}, 05:044, 2021.

\bibitem{Holz:2005df}
Daniel~E. Holz and Scott~A. Hughes.
\newblock {Using gravitational-wave standard sirens}.
\newblock {\em Astrophys. J.}, 629:15--22, 2005.

\bibitem{Chen:2017rfc}
Hsin-Yu Chen, Maya Fishbach, and Daniel~E. Holz.
\newblock {A two per cent Hubble constant measurement from standard sirens
  within five years}.
\newblock {\em Nature}, 562(7728):545--547, 2018.

\bibitem{Fishbach:2018gjp}
M.~Fishbach, R.~Gray, I.~Maga{\~n}a~Hernandez, et~al.
\newblock {A Standard Siren Measurement of the Hubble Constant from GW170817
  without the Electromagnetic Counterpart}.
\newblock {\em Astrophys. J. Lett.}, 871(1):L13, 2019.

\bibitem{Gray:2019ksv}
Rachel Gray, Ignacio~Maga{\~n}a Hernandez, Hong Qi, et~al.
\newblock {Cosmological inference using gravitational wave standard sirens: A
  mock data analysis}.
\newblock {\em Phys. Rev. D}, 101(12):122001, 2020.

\bibitem{LIGOScientific:2021aug}
R.~Abbott et~al.
\newblock {Constraints on the cosmic expansion history from GWTC-3}.
\newblock 11 2021.

\bibitem{Tamanini:2016zlh}
Nicola Tamanini, Chiara Caprini, Enrico Barausse, Alberto Sesana, Antoine
  Klein, and Antoine Petiteau.
\newblock {Science with the space-based interferometer eLISA. III: Probing the
  expansion of the Universe using gravitational wave standard sirens}.
\newblock {\em JCAP}, 04:002, 2016.

\bibitem{Yang:2022cgm}
Yuchan Yang, Wen-Biao Han, Qianyun Yun, Peng Xu, and Ziren Luo.
\newblock {Tracing astrophysical black hole seeds and primordial black holes
  with LISA-Taiji network}.
\newblock {\em Mon. Not. Roy. Astron. Soc.}, 512(4):6217--6224, 2022.

\bibitem{Alexander:2004us}
Stephon Haigh-Solom Alexander, Michael~E. Peskin, and Mohammad~M.
  Sheikh-Jabbari.
\newblock {Leptogenesis from gravity waves in models of inflation}.
\newblock {\em Phys. Rev. Lett.}, 96:081301, 2006.

\bibitem{Satoh:2007gn}
Masaki Satoh, Sugumi Kanno, and Jiro Soda.
\newblock {Circular Polarization of Primordial Gravitational Waves in
  String-inspired Inflationary Cosmology}.
\newblock {\em Phys. Rev. D}, 77:023526, 2008.

\bibitem{Obata:2014loa}
Ippei Obata, Takashi Miura, and Jiro Soda.
\newblock {Chromo-Natural Inflation in the Axiverse}.
\newblock {\em Phys. Rev. D}, 92(6):063516, 2015.
\newblock [Addendum: Phys.Rev.D 95, 109902 (2017)].

\bibitem{Adshead:2012kp}
Peter Adshead and Mark Wyman.
\newblock {Chromo-Natural Inflation: Natural inflation on a steep potential
  with classical non-Abelian gauge fields}.
\newblock {\em Phys. Rev. Lett.}, 108:261302, 2012.

\bibitem{Kahniashvili:2005qi}
Tina Kahniashvili, Grigol Gogoberidze, and Bharat Ratra.
\newblock {Polarized cosmological gravitational waves from primordial helical
  turbulence}.
\newblock {\em Phys. Rev. Lett.}, 95:151301, 2005.

\bibitem{Ellis:2020uid}
John Ellis, Malcolm Fairbairn, Marek Lewicki, Ville Vaskonen, and Alastair
  Wickens.
\newblock {Detecting circular polarisation in the stochastic gravitational-wave
  background from a first-order cosmological phase transition}.
\newblock {\em JCAP}, 10:032, 2020.

\bibitem{Seto:2006hf}
Naoki Seto.
\newblock {Prospects for direct detection of circular polarization of
  gravitational-wave background}.
\newblock {\em Phys. Rev. Lett.}, 97:151101, 2006.

\bibitem{Flanagan:1993ix}
Eanna~E. Flanagan.
\newblock {The Sensitivity of the laser interferometer gravitational wave
  observatory (LIGO) to a stochastic background, and its dependence on the
  detector orientations}.
\newblock {\em Phys. Rev. D}, 48:2389--2407, 1993.

\bibitem{Allen:1997ad}
Bruce Allen and Joseph~D. Romano.
\newblock {Detecting a stochastic background of gravitational radiation: Signal
  processing strategies and sensitivities}.
\newblock {\em Phys. Rev. D}, 59:102001, 1999.

\bibitem{Seto:2020zxw}
Naoki Seto.
\newblock {Measuring Parity Asymmetry of Gravitational Wave Backgrounds with a
  Heliocentric Detector Network in the mHz Band}.
\newblock {\em Phys. Rev. Lett.}, 125:251101, 2020.

\bibitem{Orlando:2020oko}
Giorgio Orlando, Mauro Pieroni, and Angelo Ricciardone.
\newblock {Measuring Parity Violation in the Stochastic Gravitational Wave
  Background with the LISA-Taiji network}.
\newblock {\em JCAP}, 03:069, 2021.

\bibitem{Omiya:2020fvw}
Hidetoshi Omiya and Naoki Seto.
\newblock {Searching for anomalous polarization modes of the stochastic
  gravitational wave background with LISA and Taiji}.
\newblock {\em Phys. Rev. D}, 102(8):084053, 2020.

\bibitem{Liu:2022umx}
Guo-Chin Liu and Kin-Wang Ng.
\newblock {Overlap reduction functions for a polarized stochastic
  gravitational-wave background in the Einstein Telescope-Cosmic Explorer and
  the LISA-Taiji networks}.
\newblock 10 2022.

\bibitem{Seto:2020mfd}
Naoki Seto.
\newblock {Gravitational Wave Background Search by Correlating Multiple
  Triangular Detectors in the mHz Band}.
\newblock {\em Phys. Rev. D}, 102(12):123547, 2020.

\bibitem{Wang:alternative2}
Gang Wang and Wen-Biao Han.
\newblock {Alternative LISA-TAIJI networks: Detectability of the isotropic
  stochastic gravitational wave background}.
\newblock {\em Phys. Rev. D}, 104(10):104015, 2021.

\bibitem{Prince:2002hp}
Thomas~A. Prince, Massimo Tinto, Shane~L. Larson, and J.~W. Armstrong.
\newblock {The LISA optimal sensitivity}.
\newblock {\em Phys. Rev. D}, 66:122002, 2002.

\bibitem{Liang:2021bde}
Zheng-Cheng Liang, Yi-Ming Hu, Yun Jiang, Jun Cheng, Jian-dong Zhang, and
  Jianwei Mei.
\newblock {Science with the TianQin Observatory: Preliminary results on
  stochastic gravitational-wave background}.
\newblock {\em Phys. Rev. D}, 105(2):022001, 2022.

\bibitem{Omiya:2021zif}
Hidetoshi Omiya and Naoki Seto.
\newblock {Correlation analysis for isotropic stochastic gravitational wave
  backgrounds with maximally allowed polarization degrees}.
\newblock {\em Phys. Rev. D}, 104(6):064021, 2021.

\bibitem{Bender:1997hs}
P.~L. Bender and D.~Hils.
\newblock {Confusion noise level due to galactic and extragalactic binaries}.
\newblock {\em Class. Quant. Grav.}, 14:1439--1444, 1997.

\bibitem{Hiscock:2000jn}
William~A. Hiscock, Shane~L. Larson, Joshua~R. Routzahn, and Ben Kulick.
\newblock {Low frequency gravitational waves from white dwarf MACHO binaries}.
\newblock {\em Astrophys. J. Lett.}, 540:L5--L8, 2000.

\bibitem{Nelemans:2001hp}
G.~Nelemans, L.~R. Yungelson, and Simon~F. Portegies~Zwart.
\newblock {The gravitational wave signal from the galactic disk population of
  binaries containing two compact objects}.
\newblock {\em Astron. Astrophys.}, 375:890--898, 2001.

\bibitem{Lamberts:2019nyk}
Astrid Lamberts, Sarah Blunt, Tyson~B. Littenberg, Shea Garrison-Kimmel, Thomas
  Kupfer, and Robyn~E. Sanderson.
\newblock {Predicting the LISA white dwarf binary population in the Milky Way
  with cosmological simulations}.
\newblock {\em Mon. Not. Roy. Astron. Soc.}, 490(4):5888--5903, 2019.

\bibitem{Yu:2010fq}
Shenghua Yu and C.~Simon Jeffery.
\newblock {The gravitational wave signal from diverse populations of double
  white dwarf binaries in the Galaxy}.
\newblock {\em Astron. Astrophys.}, 521:A85, 2010.

\bibitem{Nelemans:2013yg}
G.~Nelemans.
\newblock {Galactic binaries with eLISA}.
\newblock {\em ASP Conf. Ser.}, 467:27--36, 2013.

\bibitem{Korol:2017qcx}
Valeriya Korol, Elena~M. Rossi, Paul~J. Groot, Gijs Nelemans, Silvia Toonen,
  and Anthony G.~A. Brown.
\newblock {Prospects for detection of detached double white dwarf binaries with
  Gaia, LSST and LISA}.
\newblock {\em Mon. Not. Roy. Astron. Soc.}, 470(2):1894--1910, 2017.

\bibitem{Korol:2020lpq}
V.~Korol et~al.
\newblock {Populations of double white dwarfs in Milky Way satellites and their
  detectability with LISA}.
\newblock {\em Astron. Astrophys.}, 638:A153, 2020.

\bibitem{Marsh:2011yj}
T.~R. Marsh.
\newblock {Double white dwarfs and LISA}.
\newblock {\em Class. Quant. Grav.}, 28:094019, 2011.

\bibitem{Seto:2004ji}
Naoki Seto.
\newblock {Annual modulation of the galactic binary confusion noise background
  and LISA data analysis}.
\newblock {\em Phys. Rev. D}, 69:123005, 2004.

\bibitem{Barack:2004wc}
Leor Barack and Curt Cutler.
\newblock {Confusion noise from LISA capture sources}.
\newblock {\em Phys. Rev. D}, 70:122002, 2004.

\bibitem{Timpano:2005gm}
Seth~E. Timpano, Louis~J. Rubbo, and Neil~J. Cornish.
\newblock {Characterizing the galactic gravitational wave background with
  LISA}.
\newblock {\em Phys. Rev. D}, 73:122001, 2006.

\bibitem{Nissanke:2012eh}
Samaya Nissanke, Michele Vallisneri, Gijs Nelemans, and Thomas~A. Prince.
\newblock {Gravitational-wave emission from compact Galactic binaries}.
\newblock {\em Astrophys. J.}, 758:131, 2012.

\bibitem{Cornish:2017vip}
Neil Cornish and Travis Robson.
\newblock {Galactic binary science with the new LISA design}.
\newblock {\em J. Phys. Conf. Ser.}, 840(1):012024, 2017.

\bibitem{Karnesis:2021tsh}
Nikolaos Karnesis, Stanislav Babak, Mauro Pieroni, Neil Cornish, and Tyson
  Littenberg.
\newblock {Characterization of the stochastic signal originating from compact
  binary populations as measured by LISA}.
\newblock {\em Phys. Rev. D}, 104(4):043019, 2021.

\bibitem{Boileau:2021sni}
Guillaume Boileau, Astrid Lamberts, Neil~J. Cornish, and Renate Meyer.
\newblock {Spectral separation of the stochastic gravitational-wave background
  for LISA in the context of a modulated Galactic foreground}.
\newblock {\em Mon. Not. Roy. Astron. Soc.}, 508(1):803--826, 2021.
\newblock [Erratum: Mon.Not.Roy.Astron.Soc. 508, 5554--5555 (2021)].

\bibitem{Korol:2021pun}
Valeriya Korol, Na'ama Hallakoun, Silvia Toonen, and Nikolaos Karnesis.
\newblock {Observationally driven Galactic double white dwarf population for
  LISA}.
\newblock {\em Mon. Not. Roy. Astron. Soc.}, 511(4):5936--5947, 2022.

\bibitem{Liu:2023qap}
Chang Liu, Wen-Hong Ruan, and Zong-Kuan Guo.
\newblock {Confusion noise from Galactic binaries for Taiji}.
\newblock {\em Phys. Rev. D}, 107(6):064021, 2023.

\bibitem{Umstatter:2005su}
Richard Umstatter, Nelson Christensen, Martin Hendry, Renate Meyer, Vimal
  Simha, John Veitch, Sarah Vigeland, and Graham Woan.
\newblock {LISA source confusion: Identification and characterization of
  signals}.
\newblock {\em Class. Quant. Grav.}, 22:S901--S912, 2005.

\bibitem{Umstatter:2005jd}
Richard Umstatter, Nelson Christensen, Martin Hendry, Renate Meyer, Vimal
  Simha, John Veitch, Sarah Vigeland, and Graham Woan.
\newblock {Bayesian modeling of source confusion in LISA data}.
\newblock {\em Phys. Rev. D}, 72:022001, 2005.

\bibitem{Wickham:2006af}
E.~D.~L. Wickham, A.~Stroeer, and A.~Vecchio.
\newblock {A Markov chain Monte Carlo approach to the study of massive black
  hole binary systems with LISA}.
\newblock {\em Class. Quant. Grav.}, 23:S819--S828, 2006.

\bibitem{Crowder:2006eu}
Jeff Crowder and Neil Cornish.
\newblock {A Solution to the Galactic Foreground Problem for LISA}.
\newblock {\em Phys. Rev. D}, 75:043008, 2007.

\bibitem{Blaut:2009si}
Arkadiusz Blaut, Stanislav Babak, and Andrzej Krolak.
\newblock {Mock LISA Data Challenge for the galactic white dwarf binaries}.
\newblock {\em Phys. Rev. D}, 81:063008, 2010.

\bibitem{Littenberg:2011zg}
Tyson~B. Littenberg.
\newblock {A detection pipeline for galactic binaries in LISA data}.
\newblock {\em Phys. Rev. D}, 84:063009, 2011.

\bibitem{Lu:2022ywf}
Yang Lu, En-Kun Li, Yi-Ming Hu, Jian-dong Zhang, and Jianwei Mei.
\newblock {An Implementation of Galactic White Dwarf Binary Data Analysis for
  MLDC-3.1}.
\newblock {\em Res. Astron. Astrophys.}, 23(1):015022, 2023.

\bibitem{Robson:2017ayy}
Travis Robson and Neil Cornish.
\newblock {Impact of galactic foreground characterization on a global analysis
  for the LISA gravitational wave observatory}.
\newblock {\em Class. Quant. Grav.}, 34(24):244002, 2017.

\bibitem{Adams:2012qw}
Matthew~R. Adams, Neil~J. Cornish, and Tyson~B. Littenberg.
\newblock {Astrophysical Model Selection in Gravitational Wave Astronomy}.
\newblock {\em Phys. Rev. D}, 86:124032, 2012.

\bibitem{Breivik:2019oar}
Katelyn Breivik, Chiara M.~F. Mingarelli, and Shane~L. Larson.
\newblock {Constraining Galactic Structure with the LISA White Dwarf
  Foreground}.
\newblock {\em Astrophys. J.}, 901(1):4, 2020.

\bibitem{LUO2020102918}
Ziren Luo, ZongKuan Guo, Gang Jin, Yueliang Wu, and Wenrui Hu.
\newblock A brief analysis to taiji: Science and technology.
\newblock {\em Results in Physics}, 16:102918, 2020.

\bibitem{Georgousi:2022uyt}
Maria Georgousi, Nikolaos Karnesis, Valeriya Korol, Mauro Pieroni, and Nikolaos
  Stergioulas.
\newblock {Gravitational waves from double white dwarfs as probes of the milky
  way}.
\newblock {\em Mon. Not. Roy. Astron. Soc.}, 519(2):2552--2566, 2022.

\bibitem{Stroeer:2006rx}
Alexander Stroeer and A.~Vecchio.
\newblock {The LISA verification binaries}.
\newblock {\em Class. Quant. Grav.}, 23:S809--S818, 2006.

\bibitem{Shah:2014nea}
Sweta Shah and Gijs Nelemans.
\newblock {Constraining parameters of white-dwarf binaries using
  gravitational-wave and electromagnetic observations}.
\newblock {\em Astrophys. J.}, 790:161, 2014.

\bibitem{Burdge:2019hgl}
Kevin~B. Burdge et~al.
\newblock {General relativistic orbital decay in a seven-minute-orbital-period
  eclipsing binary system}.
\newblock {\em Nature}, 571(7766):528--531, 2019.

\bibitem{Burdge:2019ron}
Kevin~B. Burdge et~al.
\newblock {Orbital Decay in a 20 Minute Orbital Period Detached Binary with a
  Hydrogen Poor Low Mass White Dwarf}.
\newblock {\em Astrophys. J. Lett.}, 886(1):L12, 2019.

\bibitem{Brown:2020uvh}
Warren~R. Brown, Mukremin Kilic, A.~B\'edard, Alekzander Kosakowski, and
  P.~Bergeron.
\newblock {A 1201 s Orbital Period Detached Binary: the First Double Helium
  Core White Dwarf LISA Verification Binary}.
\newblock {\em Astrophys. J. Lett.}, 892(2):L35, 2020.

\bibitem{Kupfer:2023nqx}
Thomas Kupfer et~al.
\newblock {LISA Galactic binaries with astrometry from Gaia DR3}.
\newblock 2 2023.

\bibitem{Zhang:2022wcp}
Xue-Hao Zhang, Shao-Dong Zhao, Soumya~D. Mohanty, and Yu-Xiao Liu.
\newblock {Resolving Galactic binaries using a network of space-borne
  gravitational wave detectors}.
\newblock {\em Phys. Rev. D}, 106(10):102004, 2022.

\bibitem{Cornish:2007if}
Neil~J. Cornish and Tyson~B. Littenberg.
\newblock {Tests of Bayesian Model Selection Techniques for Gravitational Wave
  Astronomy}.
\newblock {\em Phys. Rev. D}, 76:083006, 2007.

\bibitem{Toonen:2012jj}
Silvia Toonen, Gijs Nelemans, and Simon Portegies~Zwart.
\newblock {Supernova Type Ia progenitors from merging double white dwarfs:
  Using a new population synthesis model}.
\newblock {\em Astron. Astrophys.}, 546:A70, 2012.

\bibitem{Digman:2022jmp}
Matthew~C. Digman and Neil~J. Cornish.
\newblock {LISA Gravitational Wave Sources in a Time-varying Galactic
  Stochastic Background}.
\newblock {\em Astrophys. J.}, 940(1):10, 2022.

\bibitem{Seto:2008di}
Naoki Seto.
\newblock {Detecting Planets around Compact Binaries with Gravitational Wave
  Detectors in Space}.
\newblock {\em Astrophys. J. Lett.}, 677:L55--L58, 2008.

\bibitem{Wong:2018amf}
Kaze W.~K. Wong, Emanuele Berti, William~E. Gabella, and Kelly
  Holley-Bockelmann.
\newblock {On the possibility of detecting ultrashort period exoplanets with
  LISA}.
\newblock {\em Mon. Not. Roy. Astron. Soc.}, 483(1):L33--L36, 2019.

\bibitem{Tamanini:2018cqb}
Nicola Tamanini and Camilla Danielski.
\newblock {The gravitational-wave detection of exoplanets orbiting white dwarf
  binaries using LISA}.
\newblock {\em Nature Astron.}, 3(9):858--866, 2019.

\bibitem{Danielski:2019rvt}
C.~Danielski, V.~Korol, N.~Tamanini, and E.~M. Rossi.
\newblock {Circumbinary exoplanets and brown dwarfs with the Laser
  Interferometer Space Antenna}.
\newblock {\em Astron. Astrophys.}, 632:A113, 2019.

\bibitem{Kang:2021bmp}
Yacheng Kang, Chang Liu, and Lijing Shao.
\newblock {Prospects for Detecting Exoplanets around Double White Dwarfs with
  LISA and Taiji}.
\newblock {\em Astron. J.}, 162(6):247, 2021.

\bibitem{Chen:2021sco}
Ju~Chen, Chang-Shuo Yan, You-Jun Lu, Yue-Tong Zhao, and Jun-Qiang Ge.
\newblock {On detecting stellar binary black holes via the LISA-Taiji network}.
\newblock {\em Res. Astron. Astrophys.}, 21(11):285, 2021.

\bibitem{2016PhRvL.116w1102S}
Alberto Sesana.
\newblock {Prospects for Multiband Gravitational-Wave Astronomy after
  GW150914}.
\newblock {\em Phys. Rev. Lett.}, 116(23):231102, 2016.

\bibitem{2020PhRvD.101j3027L}
Shuai Liu, Yi-Ming Hu, Jian-dong Zhang, and Jianwei Mei.
\newblock {Science with the TianQin observatory: Preliminary results on
  stellar-mass binary black holes}.
\newblock {\em Phys. Rev. D}, 101(10):103027, 2020.

\bibitem{2022RAA....22a5020C}
Ju~{Chen}, Changshuo {Yan}, Youjun {Lu}, Yuetong {Zhao}, and Junqiang {Ge}.
\newblock {On Dark Gravitational Wave Standard Sirens as Cosmological Inference
  and Forecasting the Constraint on Hubble Constant using Binary Black Holes
  Detected by Deci-hertz Observatory}.
\newblock {\em Research in Astronomy and Astrophysics}, 22(1):015020, January
  2022.

\bibitem{2021MNRAS.500.1421Z}
Yuetong {Zhao} and Youjun {Lu}.
\newblock {Stochastic Gravitational Wave Background and Eccentric Stellar
  Compact Binaries}.
\newblock {\em Mon. Not. Roy. Astron. Soc.}, 500(1):1421--1436, 2020.

\bibitem{2021arXiv211103606T}
{The LIGO Scientific Collaboration}, {the Virgo Collaboration}, and {the KAGRA
  Collaboration}.
\newblock {GWTC-3: Compact Binary Coalescences Observed by LIGO and Virgo
  During the Second Part of the Third Observing Run}.
\newblock {\em arXiv e-prints}, page arXiv:2111.03606, November 2021.

\bibitem{2021arXiv211103634T}
{The LIGO Scientific Collaboration}, {the Virgo Collaboration}, and {the KAGRA
  Collaboration}.
\newblock {The population of merging compact binaries inferred using
  gravitational waves through GWTC-3}.
\newblock {\em arXiv e-prints}, page arXiv:2111.03634, November 2021.

\bibitem{LISA2000}
{LISA~Study~Team}.
\newblock Lisa (laser interferometer space antenna): A cornerstone mission for
  the observation of gravitational waves.
\newblock Technical Report~11, ESA-SCI, 2000.
\newblock System and Technology Study Report.

\bibitem{Dhurandhar:2004rv}
S.~V. Dhurandhar, K.~Rajesh~Nayak, S.~Koshti, and J.~Y. Vinet.
\newblock {Fundamentals of the LISA stable flight formation}.
\newblock {\em Class. Quant. Grav.}, 22:481--488, 2005.

\bibitem{Klein:2015hvg}
Antoine Klein et~al.
\newblock {Science with the space-based interferometer eLISA: Supermassive
  black hole binaries}.
\newblock {\em Phys. Rev. D}, 93(2):024003, 2016.

\bibitem{ppe2009PhRvD..80l2003Y}
Nicolas Yunes and Frans Pretorius.
\newblock {Fundamental Theoretical Bias in Gravitational Wave Astrophysics and
  the Parameterized Post-Einsteinian Framework}.
\newblock {\em Phys. Rev. D}, 80:122003, 2009.

\bibitem{yunes2012PhRvD..86b2004C}
Katerina Chatziioannou, Nicolas Yunes, and Neil Cornish.
\newblock {Model-Independent Test of General Relativity: An Extended
  post-Einsteinian Framework with Complete Polarization Content}.
\newblock {\em Phys. Rev. D}, 86:022004, 2012.
\newblock [Erratum: Phys.Rev.D 95, 129901 (2017)].

\bibitem{Wang:2021mou}
Gang Wang and Wen-Biao Han.
\newblock {Observing gravitational wave polarizations with the LISA-TAIJI
  network}.
\newblock {\em Phys. Rev. D}, 103(6):064021, 2021.

\bibitem{Romano:2016dpx}
Joseph~D. Romano and Neil~J. Cornish.
\newblock {Detection methods for stochastic gravitational-wave backgrounds: a
  unified treatment}.
\newblock {\em Living Rev. Rel.}, 20(1):2, 2017.

\bibitem{Thrane:2013oya}
Eric Thrane and Joseph~D. Romano.
\newblock {Sensitivity curves for searches for gravitational-wave backgrounds}.
\newblock {\em Phys. Rev. D}, 88(12):124032, 2013.

\bibitem{BBO:2005}
Jeff Crowder and Neil~J. Cornish.
\newblock {Beyond LISA: Exploring future gravitational wave missions}.
\newblock {\em Phys. Rev. D}, 72:083005, 2005.

\bibitem{DECIGO:2006}
S.~Kawamura et~al.
\newblock {The Japanese space gravitational wave antenna DECIGO}.
\newblock {\em Class. Quant. Grav.}, 23:S125--S132, 2006.

\bibitem{Zhao:2023}
Yuetong {Zhao}, Youjun {Lu}, Changshuo {Yan}, Zhiwei {Chen}, and Wei-Tou {Ni}.
\newblock {Multiband gravitational wave observations of stellar binary black
  holes at the low to middle and high frequencies}.
\newblock {\em Monthly Notices of the Royal Astronomical Society},
  522(2):2951--2966, June 2023.

\bibitem{Ni:2012eh}
Wei-Tou Ni.
\newblock {ASTROD-GW: Overview and Progress}.
\newblock {\em Int. J. Mod. Phys. D}, 22:1341004, 2013.

\bibitem{Ni:2016wcv}
Wei-Tou Ni.
\newblock {Gravitational wave detection in space}.
\newblock {\em Int. J. Mod. Phys. D}, 25(14):1630001, 2016.

\bibitem{Baker:2019pnp}
John Baker et~al.
\newblock {Space Based Gravitational Wave Astronomy Beyond LISA}.
\newblock {\em Bull. Am. Astron. Soc.}, 51(7):243, 2019.

\bibitem{Sesana:2019vho}
Alberto Sesana et~al.
\newblock {Unveiling the gravitational universe at $\mu$-Hz frequencies}.
\newblock {\em Exper. Astron.}, 51(3):1333--1383, 2021.

\bibitem{Martens:2023mgm}
Waldemar Martens, Michael Khan, and Jean-Baptiste Bayle.
\newblock {LISAmax: Improving the Gravitational-Wave Sensitivity by Two Orders
  of Magnitude}.
\newblock 4 2023.

\bibitem{Wang:2023jct}
Gang Wang, Zhen Yan, Bin Hu, and Wei-Tou Ni.
\newblock {Investigating galactic double white dwarfs for sub-milliHz
  gravitational wave missions}.
\newblock 2 2023.

\end{thebibliography}

\end{document}